\documentclass[aps,pre,twocolumn,showpacs,floatfix,superscriptaddress, graphics]{revtex4-1}
\usepackage{amsmath,amssymb,graphicx,color,braket,hyphenat,makeidx,xcolor}
\usepackage[ocgcolorlinks,colorlinks=true,linkcolor=blue,citecolor=red,linktocpage=true]{hyperref}

\newcommand{\blue}{}

\newcommand{\tr}{\mathrm{Tr}}
\newcommand{\E}{\mathcal{E}}

\begin{document}

\title{Autonomous thermal machine for amplification and control of energetic coherence}

\author{Gonzalo Manzano}
\affiliation{Scuola Normale Superiore, Piazza dei Cavalieri 7, I-56126, Pisa, Italy.}
\affiliation{International Center for Theoretical Physics ICTP, Strada Costiera 11, I-34151, Trieste, Italy.}


\author{Ralph Silva}
\affiliation{Group D\'epartement de Physique Applique\`e, Universit\'e de Gen\`eve, 1211 Gen\`eve, Switzerland.}
\affiliation{Institute for Theoretical Physics, ETH Zurich, 8093 Zurich, Switzerland}
\author{Juan M.R. Parrondo}
\affiliation{Departamento de F\'isica At\'omica, Molecular y Nuclear and  GISC, Universidad Complutense Madrid, 28040 Madrid, Spain.}

\begin{abstract}
We present a model for an autonomous quantum thermal machine comprised of two qubits capable of manipulating and even amplifying the local coherence in a non-degenerate external system. The machine uses only thermal resources, namely, contact with two heat baths at different temperatures, 
and the external system has a non-zero initial amount of coherence. The method we propose allows for an interconversion between energy, both work and heat, and coherence in an autonomous configuration working in out-of-equilibrium conditions. This model raises interesting questions about the role of 
fundamental limitations on transformations involving coherence and opens up new possibilities in the manipulation of coherence by autonomous thermal machines.
\end{abstract}

\maketitle

\section{Introduction}

Coherence is a defining feature of quantum mechanics. The superposition principle predicts the existence of coherent (or superposition) states, in which a quantum system can be in many states with different properties at once, at difference from statistical mixtures. Coherence is responsible for interference phenomena and becomes a crucial element in most applications of quantum science \cite{Nielsen-book, Wiseman-book}. It may also play an important role in biological processes such as photosynthetic light harvesting or avian magnetoreception \cite{Lambert-biology, Gauger-avian, Collini-algae}. In addition, a rigorous abstract framework to properly quantify coherence and its interconversion in a resource-theory fashion has been developed in recent years \cite{Aberg06, Vaccaro08, Plenio-coherence, Adesso-coherence,Winter-coherence,Marvian-asymmetry,Streltsov-coherence}.

In the context of quantum thermodynamics, the role that coherence may play in boosting thermodynamic tasks such as work extraction, refrigeration or information erasure, has recently come under increasing investigation \cite{Paul-review, Anders-review, Pla17, Pat17}. Coherence allows extracting a greater amount of work from single quantum systems \cite{Paul-work, Aberg-coherence, Korzekwa-coherence, Kammerlander-coherence}, improves the performance of thermal reservoirs \cite{Scully-coherentbath, Huang-squeezing, Hai-coherentbeam, Hardal-engine, Gonzalo-squeezing, squeezing-demon}, increases power in thermal machines \cite{Scully-noisycoherence, Brunner-entanglement, Uzdin-equivalence, Brandner-feedback}, and leads to temperatures unattainable by incoherent fridges \cite{Mitchison-oscillations, Brask15}.

All those works investigate the benefits from using coherence to improve traditional thermodynamic tasks. Here we are concerned with the opposite perspective, that is, the generation of coherence from other thermodynamic resources, {\blue since it may provide new insights about the link between them}. Within this new perspective, generation of \emph{degenerate} coherence by autonomous machines \cite{Brask-entanglement} or by collective interactions with a common thermal reservoir \cite{Cakmak} has been recently considered. However, at difference from previous works, {\blue here we extend our interest to the manipulation of \emph{energetic} coherence}, i.e. coherence between states with different energies. Energetic coherence is a particularly valuable resource \cite{Streltsov-coherence, Bartlett-review}. It behaves as a quantum clock \cite{Woods-clock}, allowing the simulation of time-dependent interactions \cite{Woods-clock,Frenzel-autonomous}, and the implementation of a much larger class of thermodynamic operations \cite{Aberg-coherence, Vaccaro18} than incoherent catalysts are able to do \cite{Lostaglio-coherence,Lostaglio-bounds,Cwiklinski-limitations}.

In this paper we present an \emph{autonomous} machine capable of controlling and even \emph{amplifying} the energetic coherence of a system. The machine is one of the simplest quantum designs, comprising two qubits (see Refs. \cite{Kosloff-review, Linden-fridge}), each coupled to a bath at different temperatures, that interacts with a steady stream of qubits with a non-zero initial amount of coherence. We find that there exist regimes in which the coherence in the stream is amplified and that it is possible to control the coherence of a broad range of qubit states.

{\blue The operations performed by our machine consist} of thermalizing interactions with the baths and energy-preserving unitary transformations, which, at first sight, may seem not be able to increase the coherence of a {\blue local} system \cite{Lostaglio-bounds}. However, this would apply only for a non-degenerate global (machine plus qubit stream) setup, a condition violated as soon as resonant interactions between the machine and the qubit stream are considered. 
{\blue In fact, for such degenerate case it is useful to distinguish between two notions of coherence, usually refer to in the literature as {\em coherence} and {\em asymmetry} \cite{Streltsov-coherence, Marvian-asymmetry, Imam-asymmetry}.} 
Measures of both quantities can be respectively defined based on the relative entropy between a state and properly defined dephased states with respect to the Hamiltonian eigenbasis \cite{Streltsov-coherence}. 
A careful analysis of the two measures reveals that in degenerate situations the former can increase under energy-preserving unitary transformations and the latter, even if it is always globally conserved, it becomes sub-additive \cite{Vaccaro08, Janzing06, Gour09}. 
Notably, both of them allow for the local amplification of energetic coherence.

{\blue Summarizing, the present work shows that thermal resources, as the difference of temperature between two thermal baths, can be used to enhance a pure quantum resource, like coherence, and explores some properties of coherence when degeneracies come into play with important consequences.} 
The paper is organized as follows. In section \ref{sec:thermo} we introduce the definition of coherence based on relative entropy and the limitations to coherence growth that arise from the laws of thermodynamics. 
Some basic features of the measures of coherence and asymmetry based on relative entropy are also discussed in appendix \ref{appC}. 
The basic setup of the machine is presented in section \ref{sec:setup} and a detailed derivation of the corresponding evolution equations for the atoms and the machine is given in appendix \ref{appA}. In section \ref{sec:coherence1} we analyze the capacity of this basic setup to amplify the coherence of a single atom in the stationary regime. 
This capacity can be used to control the coherence by  a concatenation of machines, as shown in section \ref{sec:procesing}. 
A detailed analysis of this setup, its main ingredients, and their respective roles in the amplification of coherence is given  in section \ref{sec:disc}. 
Finally, in section \ref{sec:conclusion} we present our main conclusions and perspectives for further research.

\section{Thermodynamics of coherence}
\label{sec:thermo}

As already mentioned, quantum coherence has been shown to play the role of a thermodynamic resource in different contexts.  
This is not surprising since coherent states have less entropy than their corresponding dephased states, i.e., the states resulting from removing the off-diagonal terms in a given basis of the Hilbert space. 

\subsection{Measures of coherence}

Although there are different possibilities to define quantitative measures of coherence \cite{Streltsov-coherence}, the one that more naturally connects with the thermodynamic formalism is based on relative entropy. The {\em relative entropy of coherence} (REC) of a state $\rho$ with respect to a basis $\mathcal B$ of the Hilbert space, usually one of the eigenbases of the Hamiltonian $H$,  is defined as \cite{Aberg06, Plenio-coherence}
\begin{equation} \label{eq:relcoherence}
 C(\rho) \equiv S(\rho || \bar{\rho}) =  S(\bar{\rho}) - S(\rho)\geq 0,
\end{equation}
where $S(\rho || \sigma) = \tr[\rho (\log \rho - \log \sigma)]$ is the quantum relative entropy and $S(\rho)=-\tr(\rho\ln\rho)$ is the von Neumann entropy (in nats). 
The fully dephased state  
\begin{equation}\label{rhofd}
\bar{\rho} = \sum_{\ket{i}\in\mathcal B} \ket{i}\braket{i|\rho|i}\bra{i}
\end{equation}
is the state with the same diagonal elements as $\rho$, and zero non-diagonal ones in the basis $\mathcal B$. We call it fully dephased to distinguish it from partially dephased states with respect to the spectral decomposition of the operator $H$ (see below). The REC in Eq. \eqref{eq:relcoherence} is monotonic under incoherent operations, constitutes a proper measure of coherence \cite{Plenio-coherence}, and can be operationally interpreted as the distillable coherence in the state $\rho$ \cite{Winter-coherence}.

{\blue Moreover, it is worth mentioning that there exists another slightly different notion of coherence coming from the more general context of reference frames \cite{Bartlett-review}, called asymmetry under time-translations, or simply asymmetry \cite{Vaccaro08, Janzing06, Marvian-asymmetry, Imam-asymmetry}, which will be of particular importance to this work. 
A measure of asymmetry based on relative entropy can also be introduced \cite{Vaccaro08,Gour09}. However, this \emph{relative entropy of asymmetry} (REA) is defined with respect to the Hamiltonian $H$ and not a basis $\mathcal B$ \cite{Marvian-asymmetry, Imam-asymmetry}:
\begin{equation} \label{eq:relasymmetrt}
 A(\rho) \equiv  S(\rho || \tilde{\rho}) =  S(\tilde{\rho}) - S(\rho),
\end{equation}
where the partially dephased state $\tilde{\rho}$ is defined as 
\begin{equation} \label{rhopd}
\tilde{\rho} = \sum_{j} \Pi_j \rho \Pi_j,
\end{equation}
$\Pi_j$ being the projectors of the spectral decomposition of $H$. Compare this partially dephased state with the fully dephased state introduced in Eq.~\eqref{rhofd}. If the operator $H$ is non-degenerate, $\tilde{\rho} = \bar{\rho}$ and then REA and REC coincide. On the other hand, if $H$ is degenerate, the partially dephased state $\tilde{\rho}$ retains off-diagonal elements in the degenerate eigenspaces.
This is because REA is only sensible to the coherence between non-degenerate energy levels, as opposed to REC, which measures the total amount of coherence (i.e. both between degenerate and non-degenerate levels). 
Consequently, REC is never smaller than REA, $C(\rho)-A(\rho)=C(\tilde \rho)\geq 0$. Furthermore, another essential property of REA is that it is non-increasing under covariant operations with respect to the time-translation symmetry defined by $H$, 
meaning (completely positive) operations $\E$ for which $\E(e^{-i H t} \rho e^{i H t}) = e^{-i H t} \E(\rho) e^{i H t}$ \cite{Vaccaro08, Janzing06, Marvian-asymmetry, Imam-asymmetry}. 

In the following we will consider the local amplification of coherence in a system with a non-degenerate Hamiltonian, for which REC and RAC are exactly equal. Nevertheless, their differences will become important later in Sec. \ref{sec:disc}.
Also, further details about the differences between REC and REA are given in appendix \ref{appC}.}

\subsection{Second law in the presence of coherence}

The machine that we introduce in this paper works with two thermal baths at different temperatures, $T_1$ and $T_2$, and is able to control the coherence of a stream of qubits. To fix the physical interpretation, we will assume that the qubits are two-level atoms (TLA) that go through the machine in a way that will be specified in section \ref{sec:setup}. Since coherence, as measured by the relative entropy (\ref{eq:relcoherence}), is directly related to the entropy of a system, the laws of thermodynamics impose some bounds on the coherence growth of the TLA. To derive these bounds, let us start by writing down the first law of thermodynamics in a stationary regime where the state of the machine does not change:
\begin{equation} \label{eq:firstlaw}
 \dot{E}_\mathrm{a} = \dot{Q}_1 + \dot{Q}_2,
\end{equation}
where $\dot{E}_\mathrm{a}$ is the rate at which energy is transferred to the  atoms, 
and $\dot{Q}_k$ is the heat flux from reservoir $k= 1,2$ into the machine. Analogously, we can state the second law as the positivity of the rate of total entropy production in the stationary regime:
\begin{equation} \label{eq:secondlaw}
 \dot{S}_\mathrm{tot} = \dot{S}_\mathrm{a} - \beta_1 \dot{Q}_1 - \beta_2 \dot{Q}_2 ~\geq 0,
\end{equation}
where $\dot{S}_\mathrm{a}$ is the change in the von Neumann entropy of the TLA stream, and $\dot{S}_k = -\beta_k \dot{Q}_k$ for $k=1,2$ is the entropy increase (in nats) in reservoir $k$, with $\beta_k=1/k_BT_k$ the inverse temperatures. In the following we assume for convenience $\beta_1 \geq \beta_2~ (T_1 \leq T_2)$. The above Eqs.~(\ref{eq:firstlaw}) and (\ref{eq:secondlaw}) establish fundamental bounds on the performance of the machine, for any operational regime. This can be better seen if we introduce the non-equilibrium free energy of the atoms in state $\rho_{\rm a}$ with respect to the reference temperature $T_1$ as $F(\rho_{\rm a}) \equiv \tr[H_\mathrm{a} \rho_{\rm a}] - k_B T_1 S(\rho_{\rm a})$, where $H_{\rm a}$ represents the Hamiltonian of the TLA. The non-equilibrium free energy characterizes the maximum amount of work extractable from a non-equilibrium state $\rho$ with the help of a thermal reservoir \cite{Paul-work, Parrondo-information}. Using Eq.(\ref{eq:firstlaw}), the second law  (\ref{eq:secondlaw}) can be written as
\begin{equation} \label{eq:secondlaw-work}
\beta_1 \dot{F}(\rho_a) \leq (\beta_1 -\beta_2) \dot{Q}_2.
\end{equation}
Eq.~(\ref{eq:secondlaw-work}) bounds the performance of heat to work conversion in the form of non-equilibrium free energy stored in the TLA stream as $\eta \equiv \dot{F}_\mathrm{a}/\dot{Q}_2 \leq \eta_\mathrm{carnot}$, with $\eta_\mathrm{carnot} = 1- \beta_2/ \beta_1$ the Carnot efficiency.

However, the nonequilibrium free energy can be further decomposed into thermal and coherence components \cite{Janzing06, Lostaglio-coherence}. Using Eq.~(\ref{eq:relcoherence}) [or equivalently Eq. \eqref{eq:relasymmetrt}], the second law inequality (\ref{eq:secondlaw-work}) can finally be expressed as a bound on the coherence amplification of the TLA stream:
\begin{equation} \label{eq:secondlawcoherence}
\dot{C}_\mathrm{a} \leq (\beta_1 - \beta_2) \dot{Q}_2 - \beta_1 \dot{{{F}}}(\bar \rho_{\rm a}) \equiv \dot C_{\rm a}^{\rm max},
\end{equation}
{\blue where $\bar{\rho}_\mathrm{a} = \tilde{\rho}_\mathrm{a}$, since $H_\mathrm{a}$ is non-degenerate.}
Following Eq.~(\ref{eq:secondlawcoherence}), amplification of energetic coherence, $\dot{C}_\mathrm{a} \geq 0$, becomes possible by means of two sources: from the  heat  flowing from the hot to the cold bath (first term) and from a decrease of the classical free energy on the atom itself (second term). Otherwise the bound $\dot {C}_{\rm a}^{\rm max}$ becomes zero, and we have that coherence can only decrease $\dot{C}_\mathrm{a} \leq 0$. In this context, an operational interpretation for the total entropy production rate in Eq.~(\ref{eq:secondlaw}), $\dot{S}_\mathrm{tot} = \dot {C}_{\rm a}^{\rm max} - \dot{C}_\mathrm{a}$, can be given as a measure of how far we are from optimal amplification, which is only achieved under reversible, equilibrium conditions.

\section{Autonomous thermal machine}
\label{sec:machine}
In this section we introduce in detail our model of the autonomous thermal machine. We discuss the main properties of the dynamical evolution including the long-time limit where the machine reaches a steady state. Then we explore the ability of the machine to amplify the local coherence of the TLA in the steady state regime. 

\subsection{Basic setup}
\label{sec:setup}

\begin{figure}[t!]
\includegraphics[width=0.9 \linewidth]{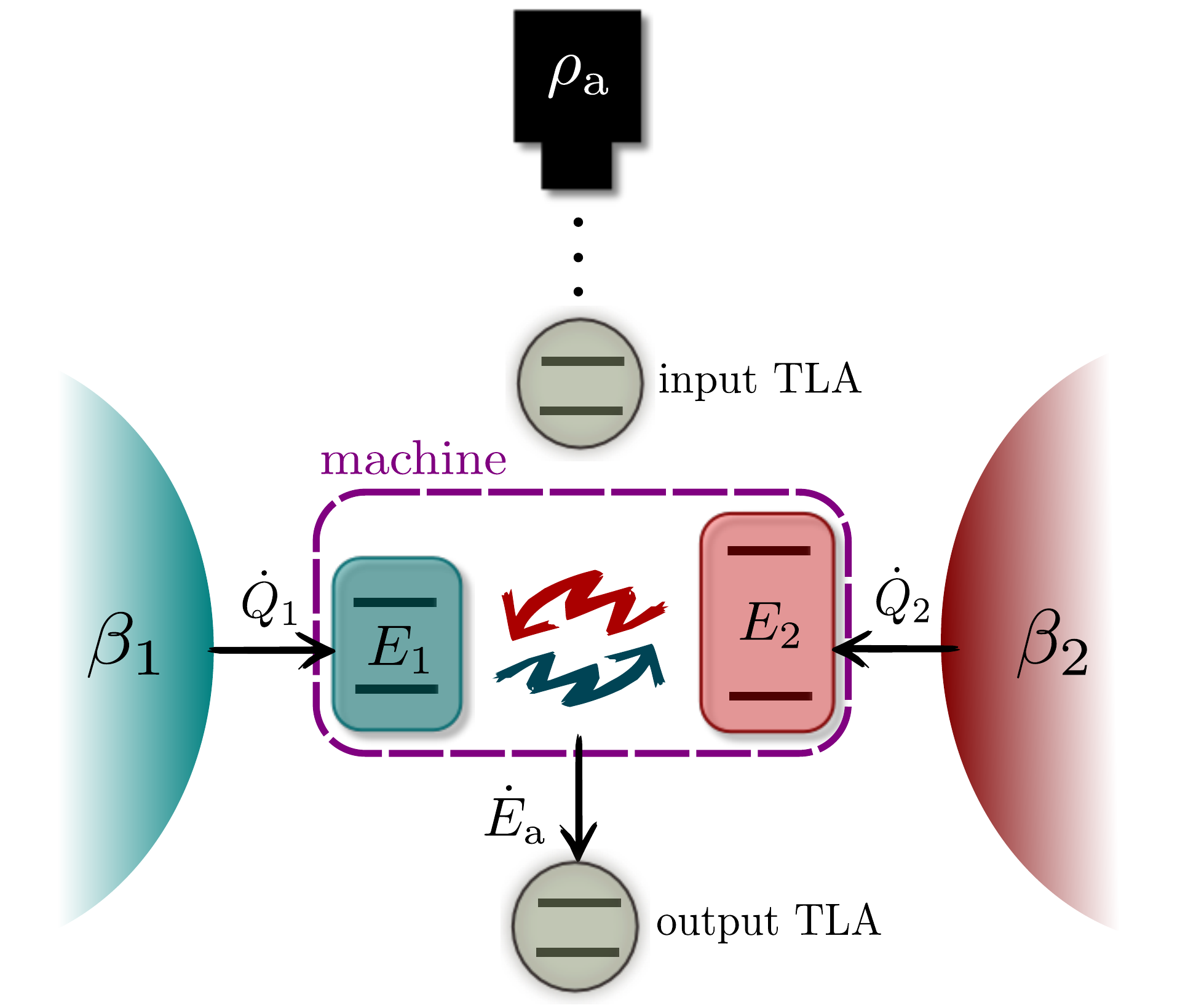}
\caption{Schematic representation of our setup: A black box throws two-level atoms (TLA) at random times in a given initial state $\rho_\mathrm{a}$. The atoms interact with the two qubits of the machine with spacings $E_1$ and $E_2$ via the energy preserving Hamiltonian $H_\mathrm{ma}$, while each qubit is coupled a to a thermal reservoir at a different temperature $(\beta_1 \geq \beta_2)$. }
\label{fig:sketch}
\end{figure}

The machine we present is sketched in Fig.~\ref{fig:sketch}  and consists of two non-interacting qubits with distinct energy spacings $E_1$ and $E_2$ (we assume for concreteness $E_2  \geq E_1$), weakly coupled to respective thermal reservoirs at different inverse temperatures, $\beta_1$ and $\beta_2$. The machine Hamiltonian is $H_{\mathrm{m}} = E_1 \sigma_1^\dagger \sigma_1^{~}  + E_2 \sigma_2^\dagger \sigma_2^{~}$, where $\sigma_1 = \ket{0}\bra{1}_{1}$ and $\sigma_2 = \ket{0}\bra{1}_{2}$ are the lowering operators of each qubit. Viewing the machine as a four level system, we can identify the middle two states $\{\ket{0}_{\rm v} \equiv \ket{1}_{1} \ket{0}_{2}, \ket{1}_{\rm v} \equiv \ket{0}_{1} \ket{1}_{2} \}$ with populations $\{p_0^{\mathrm v}, p_1^{\mathrm v} \}$ and spacing $E_2 - E_1$. We refer to this subspace as the \emph{virtual qubit} \cite{Brunner-virtual}. In the absence of any other interactions, the two qubits remain in thermal equilibrium with their respective reservoirs. In such conditions, a (virtual) inverse temperature can be ascribed to the virtual qubit via the Gibbs ratio, and reads 
\begin{equation} \label{eq:virtualtemp}
\beta_{\rm v} \equiv \frac{\ln(p_0^{\mathrm v}/p_1^{\mathrm v})}{E_2 - E_1} = \frac{\beta_2 E_2 - \beta_1 E_1}{E_2 - E_1},
\end{equation}
which can take any desired value by design. The basic idea underlying small thermal machines is to make use of the virtual qubit at a properly tuned virtual temperature to perform thermodynamic tasks (cooling, heating, storing work) upon an external system, this task is powered by the temperature difference in the reservoirs \cite{Linden-fridge, Brunner-virtual, Silva-performance}.

Together with the two-qubit machine, we introduce a third element consisting of a sequence of two-level atoms (TLA) which are sent through the machine at random times that follow Poissonian statistics with rate $r$. The atoms are all prepared in the same (but arbitrary) initial state, $\rho_\mathrm{a}$, and are assumed to interact resonantly with the virtual qubit of the machine one at a time (see Fig. \ref{fig:sketch}). The Hamiltonian of a single TLA in the sequence reads $H_{\rm a} = (E_{2} - E_{1}) \sigma_{\rm a}^\dagger \sigma_{\rm a}^{~}$, where $\sigma_{\rm a} = \ket{0}\bra{1}_{\rm a}$.

The interaction between the atom and the machine when the atom passes through is 
\begin{equation} \label{eq:Hint}
H_{\rm ma} =  \hbar g(t) (\sigma_{\rm v}^{~} \sigma_{\rm a}^\dagger + \sigma_{\rm v}^\dagger \sigma_{\rm a}^{~}) \equiv \hbar g(t) {V}, 
\end{equation}
$\sigma_{\rm v} \equiv \sigma_{1}^\dagger \sigma_2^{~} = \ket{0} \bra{1}_{\rm v}$ being the lowering operator of the virtual qubit, and $g(t)$ a time-dependent coupling strength vanishing outside the interaction region. It is convenient to define the effective strength $\phi = \int_{t_0}^{t_0 +\tau_\mathrm{i}} g(t) dt$, $\tau_\mathrm{i}$ being the interaction time and $t_0$ arbitrary. The interaction Hamiltonian $H_\mathrm{ma}$ preserves energy, i.e. $[H_\mathrm{a} + H_\mathrm{m}, H_\mathrm{ma}] = 0$, and involves a three-body interaction allowing the transfer of excitations among qubits $1$, $2$, and the TLA. 
This implies that no external sources of work are needed to make the TLA interact with the machine. Moreover, $\phi$ is taken to be the same for every TLA in the sequence. As we will shortly see, this TLA stream can act both as a passive element operated by the machine, but also as an active source driving the machine to a stationary state with non-zero coherence in its energy basis. That in turn, will result in a steady increase of the local coherence in the flying TLA.

Assuming a small interaction time $\tau_\mathrm{i}$, such that the effect of the thermal reservoirs can be neglected during the passage of the atoms, a master equation in Lindblad form can be obtained for the reduced dynamics of the machine using Born-Markov and rotating-wave approximations \cite{Wiseman-book}. On the other hand, the effect of the machine on each atom is given by a completely-positive and trace-preserving (CPTP) map $\mathcal{A}$. In the interaction picture with respect to $H_\mathrm{m} + H_\mathrm{a}$ they read (see appendix \ref{appA} for details)
\begin{subequations} \label{eq:masters}
\begin{align} \label{eq:mastermachine}
&\dot{\rho}_{\mathrm{m}} = -i r \phi [ V_\mathrm{m}, \rho_\mathrm{m}] + \sum_{k = \mathrm{v}, 1, 2,} \mathcal{D}_{k}(\rho_\mathrm{m}) \equiv \mathcal{L}_\mathrm{m}(\rho_\mathrm{m}), ~~\\
\label{eq:masteratoms} 
&\mathcal{A}({\rho}_{\mathrm{a}}) = \rho_\mathrm{a} -i \phi [ V_\mathrm{a}, \rho_\mathrm{a}] + \mathcal{D}_{\mathrm a}(\rho_\mathrm{a}),
\end{align}
\end{subequations}
where the coherent (driving-field like) terms read $V_\mathrm{m} = \tr_\mathrm{a}[V \rho_\mathrm{a}] = \sigma_{\rm v}^{~} \braket{\sigma_{\rm a}^\dagger} + \sigma_{\rm v}^\dagger \braket{\sigma_{\rm a}^{~}}$ in Eq.~(\ref{eq:mastermachine}), and analogously $V_\mathrm{a} = \tr_\mathrm{m}[V \rho_\mathrm{m}(t)]$ in Eq.~(\ref{eq:masteratoms}), whose strengths depend on the off-diagonal elements (in the energy eigenbasis) of $\rho_{\rm a}$ and $\rho_\mathrm{m}(t)$ respectively. In addition, we obtain the following dissipators that account for the energy jumps induced by both the interaction and the thermal reservoirs:
\begin{align}\label{eq:dissipators}
 \mathcal{D}_{k}(\rho) =&~ \gamma_\downarrow^k \left(\sigma_k^{~} \rho \sigma_k^\dagger - \frac{1}{2}\{\sigma_k^\dagger \sigma_k^{~}, \rho \} \right) \nonumber \\ 
 &+ \gamma_\uparrow^k \left(\sigma_k^\dagger \rho \sigma_k^{~} - \frac{1}{2}\{\sigma_k^{~} \sigma_k^\dagger, \rho \} \right),
\end{align}
with $k= 1, 2, \mathrm{v}, \mathrm{a}$. Here the rates of emission and absorption processes induced by the thermal reservoirs 
obey detailed balance $\gamma_\downarrow^k = \gamma_\uparrow^k~ e^{\beta_k E_k}$ for $k=1,2$, and we have the following rates 
from machine-atom interactions $\gamma_\downarrow^{\rm v} = r \phi^2 \braket{\sigma_{\rm a}^{~} \sigma_{\rm a}^\dagger}$ and $
\gamma_\uparrow^{\rm v} = r \phi^2 \braket{\sigma_{\rm a}^{\dagger} \sigma_{\rm a}^{~}}$ for $\mathcal{D}_\mathrm{v}$, together 
with $\gamma_\downarrow^{\rm a}(t) = \phi^2 \braket{\sigma_{\rm v}^{~} \sigma_{\rm v}^\dagger}_t$ and $\gamma_\uparrow^{\rm a}(t) 
= \phi^2 \braket{\sigma_{\rm v}^{\dagger} \sigma_{\rm v}^{~}}_t$ for $\mathcal{D}_\mathrm{a}$. Notice that the dynamics of the TLA, 
in contrast to the machine dynamics, is characterized by time-dependent coefficients, $\gamma_{\uparrow \downarrow}^{\mathrm a}
(t) \geq 0 ~\forall t$.
Self-consistency of Eqs. (\ref{eq:mastermachine}) and (\ref{eq:masteratoms}) requires $\tau_\mathrm{i} \ll 1/\gamma_0^k$  with $\gamma_0^k \equiv \gamma_\downarrow^k - \gamma_\uparrow^k$, together with $\phi^2 \ll E_2 - E_1$ and $\gamma_0^k \ll E_k$, $k = 1, 2$ (see App. \ref{appAmaster}).

Importantly, the interplay of coherent and dissipative terms in Eq. (\ref{eq:mastermachine}) implies that in the long-time run, when sufficiently many atoms have interacted with the machine, the latter  reaches  a steady state, $\mathcal{L}_\mathrm{m}(\pi_\mathrm{m}) = 0$, that has non-zero coherence in the virtual qubit. This state can be obtained analytically, but it shows a complicated dependence on the initial preparation of the TLA and all other parameters.
{\blue While the machine is interesting to investigate, for example to ascertain whether some non-zero entanglement can be maintained between the two machine qubits, in this work we focus on the effect on the TLA stream.}

The dynamics of the TLA stream is obtained by inserting $\pi_\mathrm{m}$ in the expectation values appearing in Eqs.~(\ref{eq:masteratoms}) and (\ref{eq:dissipators}). Once the machine is in a steady state, all the output atoms reach the same state after interacting with the machine, with only an infinitesimal change to their initial state $\rho_\mathrm{a}$ (since $\phi$ is small). However, dynamical control and finite state transformations over individual atoms can be achieved in the extended configuration considered in Sec.~\ref{sec:procesing}.

\subsection{Coherence amplification}
\label{sec:coherence1}

For the machine working at steady state conditions, $\dot{C}_\mathrm{a}$ and $\dot{C}_{\rm a}^{\rm max}$ can be computed analytically (see appendix \ref{appB}). Recall that the maximum coherence growth rate $\dot {C}_{\rm a}^{\rm max}$ is given in terms of the free energy and the heat flows by Eq.~(\ref{eq:secondlawcoherence}). 
In the stationary regime, the energy change of the TLA stream and its change in von Neumann entropy are given respectively by  
\begin{align} \label{eq:Ea}
\dot{E}_\mathrm{a} &= r ~\tr[H_\mathrm{a} (\mathcal{A}(\rho_\mathrm{a}) -\rho_\mathrm{a})], \\ \label{Eq:Sa}
\dot{S}_\mathrm{a} &= - r ~\tr[\mathcal{A}(\rho_\mathrm{a}) \ln \mathcal{A}(\rho_\mathrm{a}) - \rho_\mathrm{a} \ln \rho_\mathrm{a}].
\end{align}
The heat flux from reservoir $k= 1,2$ reads 
\begin{equation}
\dot{Q}_k = \tr[H_\mathrm{m} \mathcal{D}_k(\rho_{\rm m})], 
\end{equation}
while $\dot{S}_k = -\beta_k \dot{Q}_k$ for $k=1,2$, is the entropy increase in reservoir $k$. 

\begin{figure}[t!]
\includegraphics[width=1.0 \linewidth]{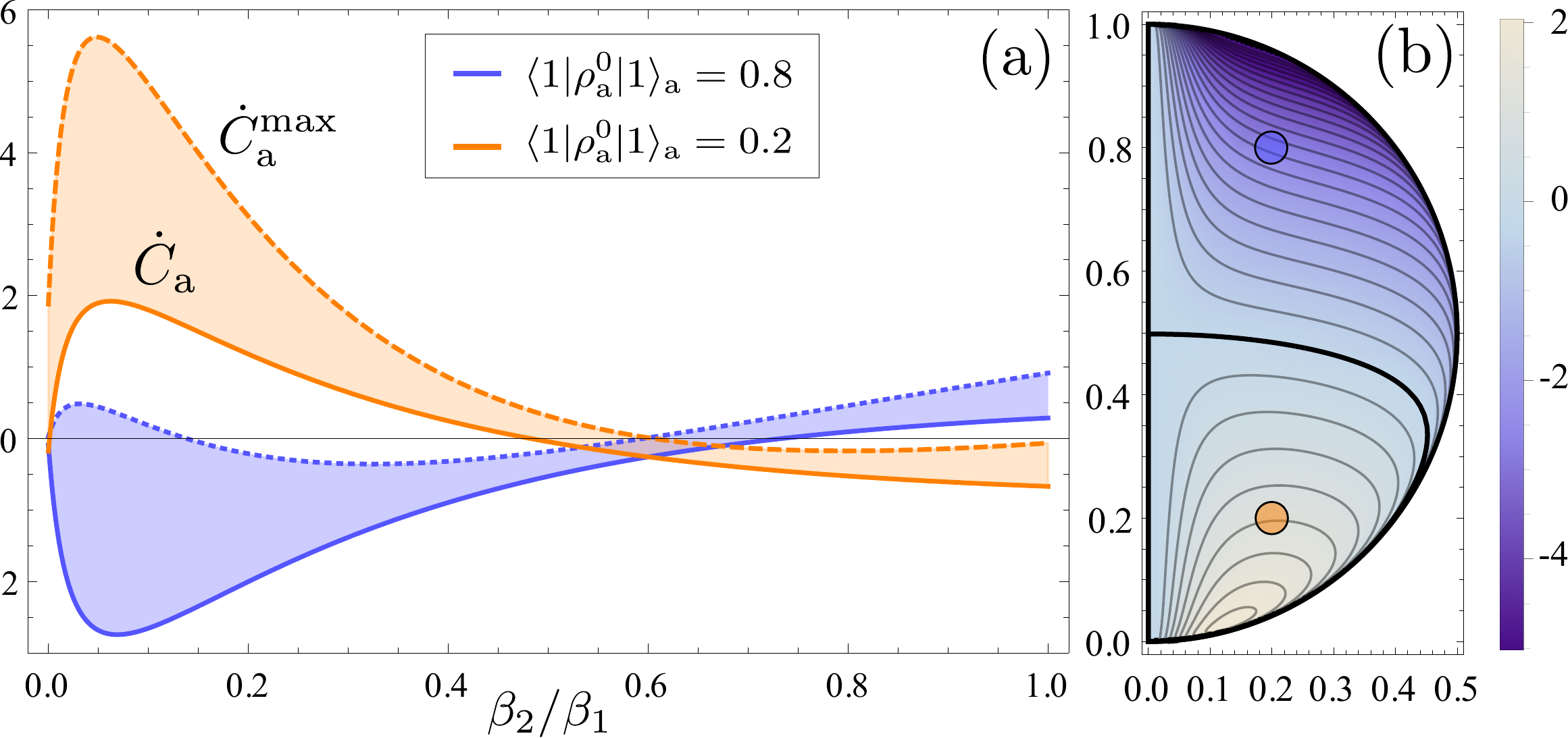}
\caption{(a) Amplification of local coherence as measured by the REC, $\dot{C}_\mathrm{a}$ (solid curves) and the bound $\dot{C}_{\rm a}^{\rm max}$ (dashed and dotted curves) as a function of $\beta_2/\beta_1$. 
The two colors represent two choices of the TLA initial state (see legend) with same initial coherence $|\bra{0} \rho_{\rm a} \ket{1}_\mathrm{a}| = 0.2$.(b) Dependence of $\dot{C}_\mathrm{a}$ on the initial preparation of the atoms for fixed $\beta_2 = 0.2 \beta_1$, displayed as a contour plot over a cross-section of the TLA Bloch sphere (in the rotating frame). 
$\dot{C}_\mathrm{a}$ and $\dot{C}_{\rm a}^{\rm max}$ are given in units of $\phi^2$ and we used $E_1 = 1.5, E_2 = 2.5$. In both plots we set $\beta_1 = 1.2, \gamma_0^k = \gamma_\downarrow^k - \gamma_\uparrow^k = 0.0025$ for $k=1,2$, $r = 2.0$ and $\phi = 0.02$.}
\label{fig:coherence}
\end{figure}

We find that coherence amplification in crossing TLA becomes possible for a broad range of initial states of the atoms and machine parameters. In Fig. \ref{fig:coherence}(a) we show $\dot{C}_\mathrm{a}$ and $\dot {C}_{\rm a}^{\rm max}$ when the reservoirs temperature ratio $\beta_2/\beta_1$ is varied. We use two paradigmatic initial states for the atom stream lying at the south (dark orange) and north (light blue) hemispheres of the Bloch sphere as depicted by the two small circles in Fig. \ref{fig:coherence}(b). In the first case we find that thermal amplification of coherence is achieved when increasing the difference of temperatures between the reservoirs until the high temperature limit, $\beta_2 E_2 \ll 1$ is approached. On the contrary, the second case illustrates the regime in which coherence is amplified at the cost of reducing classical non-equilibrium free energy of the atoms. Notice that this process can occur in the limit $\beta_2 \rightarrow \beta_1$, that is, it does not need any input power from the machine. Optimal amplification $\dot{C}_{\rm a}^{\rm max}$  cannot be achieved in any case, the shaded regions highlighting the total entropy production rate in the setup. In this context is interesting to notice the point $\beta_2 \rightarrow 0.6 \beta_1/$, where $\dot{C}_\mathrm{a}^\mathrm{max}$ becomes zero, and, consequently, entropy production is entirely due to decoherence processes. In Fig. \ref{fig:coherence}(b), the contour lines show the dependence of $\dot{C}_\mathrm{a}$ on the initial state of the input atoms in the sequence, $\rho_\mathrm{a}$, for a given difference of temperatures. There the black thick contour corresponds to $\dot{C}_\mathrm{a} = 0$. We can appreciate that coherence amplification becomes possible for a broad range of initial states with non-zero initial 
coherence inside the south hemisphere of the atoms Bloch sphere.

{\blue The physical mechanism underlying coherence amplification in our machine can be understood by splitting its operation in steady state conditions into two steps. In the first step an incoming TLA in state $\rho_\mathrm{a}$ interacts with the virtual qubit of the machine in state $\pi_\mathrm{m}$, through the interaction $H_\mathrm{ma}$ for some small amount of time $\tau_\mathrm{i}$ [Eqs. \eqref{eq:unitary} and \eqref{eq:totaleffect} in App. \ref{appA}]. During this unitary evolution, and thanks to the degeneracy in the global machine-TLA system provided by $[H_\mathrm{ma}, H_\mathrm{m} + H_\mathrm{a}] = 0$, both the TLA and the virtual qubit may increase their local coherences. This is the case when both initial states of the virtual qubit and the incoming TLA have some initial amount of coherence and either one or the other show population inversion (a proof is given in App. \ref{appC4}). 
This is in accordance with our general result in Eq. (\ref{eq:secondlawcoherence}), from which we learn that amplification of coherence requires either a heat flow between two different temperatures, or the release of (diagonal) free energy by the TLA itself. Then, in the second step, the machine qubits interact with their respective thermal reservoirs at different temperatures for some (small) amount of time, until the state $\pi_\mathrm{m}$ of the machine is recovered. In this second process some of the coherence in the virtual qubit of the machine is lost in the reservoirs, but its population bias is recovered, and the next interaction can take place.}

\section{Coherence processing} \label{sec:procesing}

\begin{figure}[t!]
\includegraphics[width=0.9 \linewidth]{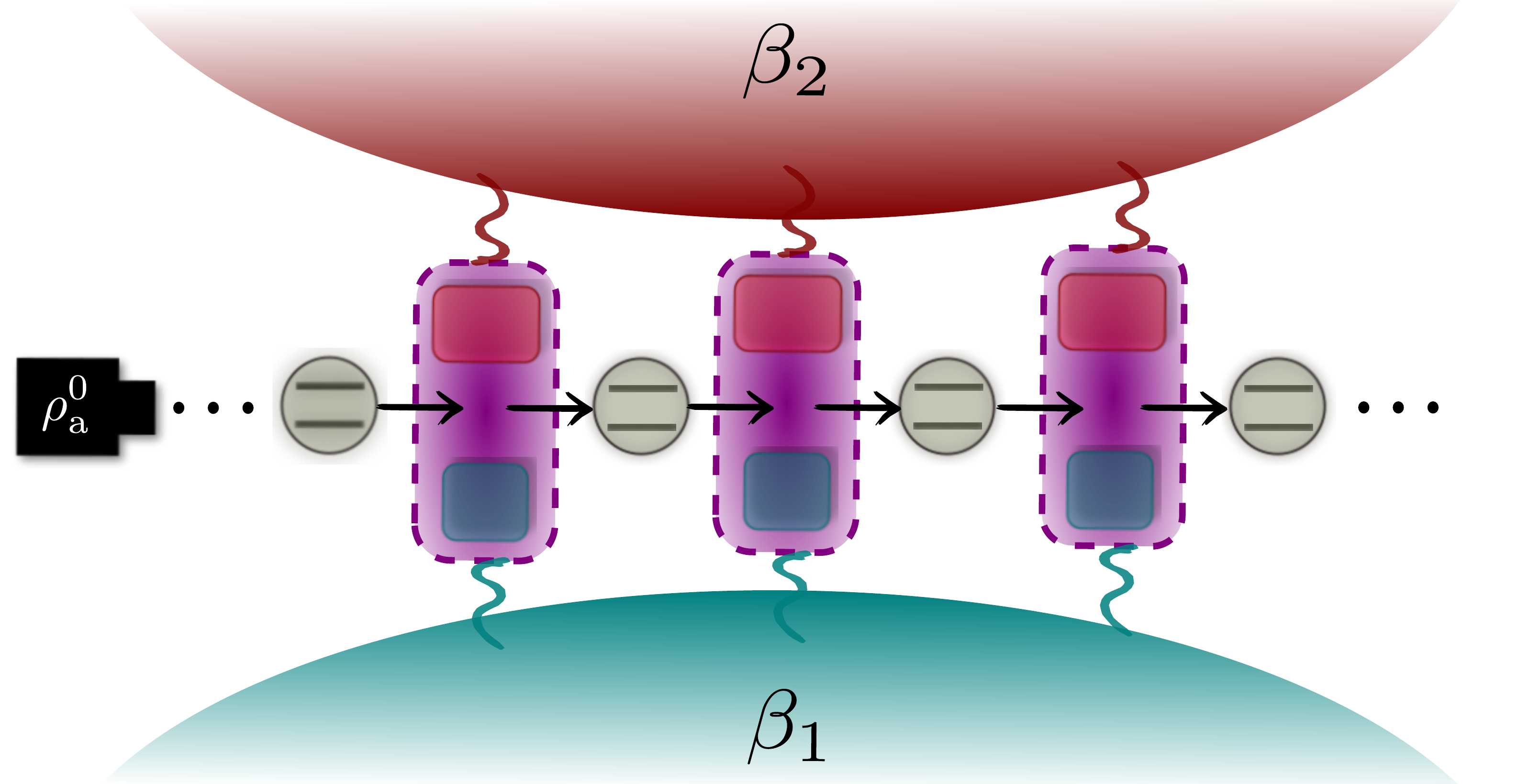}
\caption{Schematic representation of the extended setup. The TLA stream crosses an array of many independent and equivalent thermal machines coupled to the same reservoirs, each of them in a different steady state.}
\label{fig:sketch2}
\end{figure}

{\blue So far our analysis of local coherence amplification applied to the ensemble of output atoms in Fig. \ref{fig:sketch}, but whose individual states change only infinitesimally (changes in REC of order $\phi^2$)}. In the following we show that the coherence of individual atoms in the sequence can be increased by a {\blue quite substantial} amount as well. 

This is accomplished in the extended configuration sketched in Fig. \ref{fig:sketch2}, where an array of thermal machines such as the one introduced above is arranged in sequence. There all the atoms are prepared in the same initial state $\rho_\mathrm{a}^0$, but each machine will now meet the atoms in a different state, as it depends on their prior interaction with previous machines. 
Nevertheless, we notice that after a sufficient time, \emph{every} machine in the sequence will reach a (different) steady state. This can be seen from the fact that the first machine in the sequence follows Eq. (\ref{eq:mastermachine}), and  after interacting with sufficiently many atoms, will reach the steady state $\pi_\mathrm{m}(\rho_\mathrm{a}^0)$ as before. After that time, the first machine induces the same dynamics on every subsequent atom and, as a consequence, the input atoms for the second machine will always be in the same state, say $\rho_\mathrm{a}^1$. The dynamics of the second machine then will be given by Eq. (\ref{eq:mastermachine}), on replacing $\rho_\mathrm{a}^0$ by $\rho_\mathrm{a}^1$. This induces the steady state $\pi_\mathrm{m}(\rho_\mathrm{a}^1)$ in the second machine and, after that, all output atoms will analogously be in a fixed state $\rho_\mathrm{a}^2$. This argument extends to the entire sequence of machines. 

When all the machines reach their steady states, then the transformation of a single TLA crossing the sequence will be given by a concatenation of CPTP maps such as the one given in Eq. (\ref{eq:masteratoms}). After crossing $n$ machines it reads
\begin{equation}\label{eq:mapstream}
 \rho_\mathrm{a}^{n} = \mathcal{A}_n \circ \mathcal{A}_{n-1} \circ \cdots \circ \mathcal{A}_{i} \circ  \cdots \circ \mathcal{A}_1 (\rho_\mathrm{a}^0)
\end{equation}
with the expectation values appearing in the $i$th map calculated for $\pi_\mathrm{m}(\rho_{\mathrm a}^{i-1})$, with $i = 1, ..., n$, {\blue that is, $V_\mathrm{a}$ and the rates $\gamma_{\downarrow \uparrow}^{\mathrm{a}}$  in Eqs. \eqref{eq:masteratoms} and \eqref{eq:dissipators}. 
Analogously, we may apply Eqs. \eqref{eq:Ea} and \eqref{Eq:Sa} (divided by $r$) for any map $\mathcal{A}_i$ in the sequence.
For a more detailed description of the setup and justification of Eq. \eqref{eq:mapstream} see App. \ref{appAext}. }

\begin{figure}[t!]
\includegraphics[width=1.0 \linewidth]{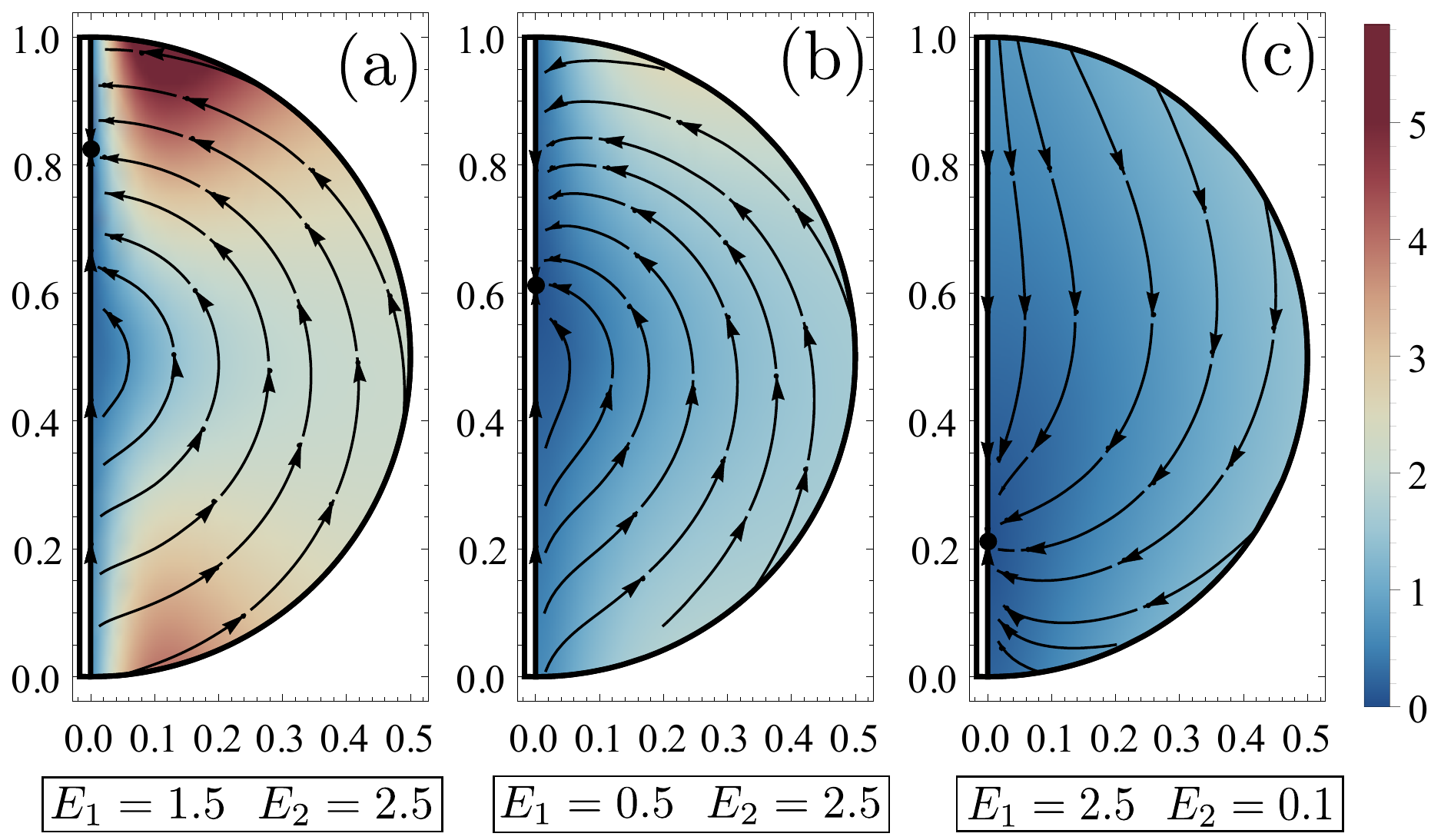}
\caption{Trajectories in a section of the Bloch sphere of individual TLA when sent through a sequence of machines ($\beta_2 = 0.1 \beta_1$). The color scale shows the modulus of the kick (in $\phi^2$ units) produced on the atom state when it crosses a machine in a given state. Set (c) has been obtained by interchanging the role of the qubits (or equivalently exchanging the temperatures of the reservoirs). 
The zero-coherence fixed points of the dynamics are depicted by the small dark circles. In all plots $\beta_1 = 1.2, \gamma_0^k = \gamma_\downarrow^k - \gamma_\uparrow^k = 0.0025$ for $k=1,2$, $r = 2.0$ and $\phi = 0.02$.}
\label{fig:coherence2}
\end{figure}

{\blue 
In Fig. \ref{fig:coherence2} we plot sample trajectories of the states followed by a single TLA when crossing the array of thermal machines on its Bloch sphere. 
The trajectories correspond to states depicted in the interaction picture, namely in a rotating frame with respect to the $z$-axis at frequency $(E_2 - E_1)/\hbar$. 
We show three different sets of trajectories (a)-(c) corresponding to different values of the machine qubits spacings $E_2$ and $E_1$. 
In any case we obtain a dissipative evolution towards an incoherent thermal steady state $\pi_\mathrm{a} = e^{-\beta_\mathrm{v} H_\mathrm{a}}/Z_\mathrm{a}$, with $Z_\mathrm{a} = \tr[e^{-\beta_{\mathrm v} H_{\mathrm a}}]$, fulfilling $\mathcal{A}_n(\pi_{\mathrm{a}})= \pi_{\mathrm{a}}$. 
This state, depicted by black dots over the $z$-axis, is reached in the limit of a large number of machines in the array, $n \rightarrow \infty$, and $\beta_\mathrm{v}$ is the virtual temperature introduced in Eq. \eqref{eq:virtualtemp}. 
}

For initial incoherent states (vertical axis) the trajectories stay always incoherent, i.e. coherence cannot be generated in the TLA if it is initially absent. 
However, we see that there exist a broad range of initial states with non-zero initial coherence for which the coherence can be amplified during the evolution. 
Even if the incoherent steady state $\pi_\mathrm{a}$ is reached when a large array of machines is considered, by preparing arrays of a finite tuned size, one can stop the trajectories at a particular target point. 
Furthermore we find that tuning $\beta_{\mathrm{v}}$ is possible by choosing the design parameters of the machine (i.e. the energies $E_1$ and $E_2$) [see Eq. \eqref{eq:virtualtemp}]. 
This allows to obtain sets of trajectories where the coherence can be amplified while also cooling the TLA, see Fig. \ref{fig:coherence2}(c). 

Importantly, we notice that the temperature difference plays a fundamental role here, enlarging the set of trajectories which can be generated, and hence increasing our ability to reach target states.

\section{Discussion}
\label{sec:disc}

The possibility of a steady increase of the local coherence of the TLA in the stationary regime of the machine that we have presented in this paper may be at first sight surprising. 
The whole dynamical evolution consists of energy-preserving unitary steps occurring at random times when a single atom interact, one at a time, with the machine, and the subsequent thermalization of the machine qubits with their respective thermal baths. 
All these transformations are special cases of thermal operations, that is, maps resulting from an interaction between a system and a thermal bath that allows the exchange of energy between the two conserving the global energy \cite{Paul-review}. 
{\blue Therefore, since it is well known that thermal operations (and indeed any phase covariant operation) cannot increase asymmetry \cite{Vaccaro08, Janzing06, Lostaglio-coherence, Lostaglio-bounds, Cwiklinski-limitations}, 
one may wonder whether the amplification of coherence in the TLA stream of our setup contradicts this or other general statements? 

\subsection{Local versus global asymmetry and coherence}

The answer to the question above is negative. The apparent contradiction relies in the fact that for a global degenerate system, as is our machine-atom setup, the non-increasing statement only applies to the asymmetry of the \emph{global} system and not necessarily to the sum of \emph{local} asymmetries (or coherences) of non-degenerate subsystems.
Notably, this is at difference from the case of non-degenerate global systems, where both the asymmetry of the global system and the sum of local asymmetries are non-increasing. 
Indeed, one of the key properties of the REA in Eq.\eqref{eq:relasymmetrt} for degenerate systems which spotlights this effect is that it becomes sub-additive. That is, for a bipartite system where the total energy is degenerate, the REA can increase just by considering the two subsystems as separate entities, 
even if the global state is uncorrelated, i.e., $A(\rho_{\rm a}\otimes\rho_{\rm m})\leq A(\rho_{\rm a})+A(\rho_{\rm m})$. The reason is that the partially dephased state $\tilde{\rho}$ can create spurious correlations between the two systems even if $\rho$ is a product uncorrelated state (see App. \ref{appC2} for a detailed discussion).
The sub-additivity property of asymmetry has been noticed before e.g. in Refs.~\cite{Vaccaro08, Janzing06, Gour09}. Indeed in Ref. \cite{Gour09}, among other results, it has been shown that $A(\rho^{\otimes N}) \leq 2 (d-1) \ln N \ll N A(\rho)$ in the asymptotic limit $N \rightarrow \infty$, $d$ being the dimension of the Hilbert space. 
Nevertheless, the consequences of sub-adittivity of asymmetry on local amplification of coherence have never been  discussed before, to the best of our knowledge. 

In any case, we have that the REA of the machine plus a single TLA state is conserved during their unitary interaction due to $[U, H_\mathrm{m} + H_\mathrm{a}] = [H_\mathrm{ma}, H_\mathrm{m} + H_\mathrm{a}] = 0$, as it corresponds to a covariant operation with respect to time-translation symmetry. 
In App.~\ref{appC3}, we obtain the following relation between the increase in local coherences and the specific structure of correlations generated in the global system [see Eq.~\eqref{applocals}]:
\begin{equation} \label{eq:localincrease}
\Delta C_\mathrm{a} + \Delta C_\mathrm{m} = I(\tilde{\rho}_{\mathrm{m a}}^\prime) - I(\tilde{\rho}_{\mathrm{m a}}) - I(\rho_\mathrm{m a}^\prime).
\end{equation}
Here $\Delta C_{i} = C (\rho_{i}^\prime) - C (\rho_{i}^{~})$, $i={\rm a,m}$, is the increase in the local REC (or, equivalently, local REA) of the TLA (machine), $\rho_\mathrm{m a}^\prime = U \rho_\mathrm{m a} U^\dagger$, with $\rho_\mathrm{m a} = \pi_\mathrm{m} \otimes \rho_\mathrm{a}$, is the global state after interaction with marginals $\rho_\mathrm{a}^\prime$ and $\rho_\mathrm{m}^\prime$, and $I(\rho_{\rm ma})$ is the quantum mutual information of the global state $\rho_{\rm ma}$. 
From Eq. \eqref{eq:localincrease} we see that local coherences can be increased when $I(\tilde{\rho}_{\mathrm{m a}}^\prime) \geq I(\tilde{\rho}_{\mathrm{m a}}) + I(\rho_\mathrm{m a}^\prime)$. This is indeed possible, as we show  in App. \ref{appC3} analyzing a specific example.

On the other hand, it is also instructive to look at the behavior of the REC, $C(\rho_\mathrm{m a})$, in the global system. As opposed to REA, REC is additive with respect to an arbitrary local basis even for degenerate systems, 
i.e. $C(\rho_{\rm a}\otimes\rho_{\rm m})= C(\rho_{\rm a})+C(\rho_{\rm m})$. However, contrary to the REA, it turns out that in the degenerate case, the REC can increase under thermal operations (see App. \ref{appC3}), providing us an alternative 
way of visualizing the local amplification effect reported in this paper (see Fig.~\ref{fig:coherence_vrs_asym}). However, we recall again that the Hamiltonians of the machine $H_{\rm m}$ and the atom $H_{\rm a}$ are non-degenerate and 
therefore it is irrelevant which measure of coherence, REC or REA, is used for the reduced states. 
}

\subsection{Ingredients for local amplification of coherence}

Moreover, in a bipartite system consisting of two qubits, it is possible to prove that the global REC increases during an energy-preserving unitary evolution only if there is a population inversion in one of the two qubits, i.e., if the excited state is more populated than the ground state (see App.\ref{appC4} for a detailed proof). 
This is a further important result since it provides a rigorous link between the amplification of local energetic coherence and population inversion, which requires either work or some other thermodynamic resource such as a temperature gradient.
 
With all this in mind, we can now distinguish the three key ingredients that allow our machine to increase the REC of the global system (and hence the local coherence in single TLA) in the stationary regime. 
To do that, it is more convenient to focus on the REC, instead of REA. The first ingredient is an energy-preserving interaction between the atom and the machine that increases the total REC of the system, similar to the unitary evolution discussed in appendix \ref{appC}. 
{\blue The condition for this to work is that either the machine or the TLA start the interaction in a state with population inversion with respect to the other one, and both of them have some non-zero initial coherence. 
Therefore the second key ingredient is the population inversion, which may be provided either directly on the initial state of the TLA, or indirectly as a temperature difference between the two baths.} 
We notice that the two thermal baths play also the role of resetting the machine to the proper state in the stationary regime. But, to obtain a state with non-zero coherence in the virtual qubit of the machine, it is necessary that the thermal relaxation is not complete. 
This is the third ingredient: a partial thermal relaxation, which is achieved by sending the atoms at a rate $r$ sufficiently high to prevent the total relaxation of the machine qubits. 

{\blue
\subsection{Catalysis and correlations}

Our mechanism for amplification of coherence is related to the one proposed by {\AA}berg in Ref. \cite{Aberg-coherence}, where a system with a high degree of coherence (a coherent battery) is used to induce arbitrary local unitaries on an external system of interest. 
We may interpret our machine as an autonomous and dissipative version of {\AA}berg's coherent battery acting on the TLA stream.
Nevertheless, we would like to point two further important differences between our setup and Aberg's one. First, our configuration allows for a steady state in the machine, $\pi_\mathrm{m}$, whereas {\AA}berg's coherent battery do not return back to its initial state after operation (see also the discussion about this point in Ref. \cite{Korzekwa-coherence}). 
Second, and more important, the increase of local coherence in each TLA in our configuration does not need an equivalent reduction of local coherence in any other system. This is a consequence of the resonant interaction between the machine and the
TLA stream, which drives the machine to a steady state and simultaneously increases the local REC and REA, as explained above.

It is also worth pointing out that, due to the repeated interaction scheme, the correlations continuously generated between the machine and the TLA may result in the generation of correlations between output atoms. This effect has been indeed recently reported for a simpler thermalization collisional model \cite{Cusumano18}. Also, very recent results concerning the possibility of distillation of coherence in a quantum thermodynamical framework \cite{Marvian14} suggest that the local amplification reported here must be accompanied by the generation of such correlations. 

These correlations have been discussed 
 by Vaccaro {\em et al} \cite{Vaccaro18} in the context of the {\AA}berg scheme, concluding that coherence is a finite resource and cannot be catalytic. The same argument applies to our machine. We would like to stress again that the REA of the global system cannot increase, and that our setup can only enhance the local REA's (or REC's) of the TLA.
However, the increase of these local coherences is still relevant if one is interested in the atoms as single and separate identities, and subsequently does not use any collective protocol or operation on the resulting TLA stream. 
In such case the correlations mentioned above are irrelevant and may be neglected. It is indeed an open problem to assess howmuch of the global REA is due to correlations, since the REA is sub-additive even for uncorrelated states, as we show in App.~\ref{appC2}.


%
%

}

\section{Summary and conclusions}
\label{sec:conclusion}

We have discussed various aspects of the thermodynamic limitations emerging when considering the interconversion between energy and coherence, and presented an autonomous thermal machine able to amplify energetic coherence using thermal resources (two thermal reservoirs at different temperatures). 
In particular, we have identified the two main thermodynamic resources for coherence generation in the setup: the spontaneous heat flow from a hot to a cold reservoir, and the reduction of the classical free energy in the system in which coherence is amplified. The interplay between these two sources 
is related to the irreversibility of the amplification process, which we characterized through the entropy production. Then we have shown how our thermal machine is able to work in nonequilibrium steady state conditions profiting from these two aforementioned resources. 

We have also identified the three key elements present in our scheme enabling coherence amplification: a unitary transformation that increases the local coherences of a degenerate bipartite system, a partial thermal relaxation, and a temperature difference that resets the machine to a state with coherence and population inversion. 
Indeed, partial thermal relaxation is a very basic idea that could have more applications in quantum thermodynamics, since it makes use of thermodynamic resources, in our case the temperature difference between the two baths, while keeping genuine quantum features like coherence. 

Interestingly, our results show that when multiple copies of an initial state with some (even if negligible) amount of coherence are allowed, a dissipative coherent catalyzer \cite{Aberg-coherence, Vaccaro18} can be created (the virtual qubit of the machine) just using energy preserving interactions between resonant transitions. 
This can be used for the coherent manipulation of qubit states (the TLA) in an extended configuration using an array of autonomous machines. 
{\blue Nonetheless, we must point that the kind of catalysis proposed here is both local and limited by dissipation. It is local because it works only for single copies of the TLA, not being allowed collective operations over the output atoms, which might not be indipendent between them. 
Furthermore, it is limited in the sense that dissipative effects prevent us from reaching arbitrary states of the TLA. }
A further  open question left concerns the possibility of combining \emph{different} machines in the same array, e.g. each of them with different spacings $E_2$ and $E_1$ in the qubits, in order to enlarge the set of reachable target states from a given initial state $\rho_\mathrm{a}^0$. 

{\blue
Moreover, we discussed some connections between our results for the autonomous manipulation of local coherence, and existing resource theories of asymmetry and coherence \cite{Vaccaro08, Janzing06, Streltsov-coherence,Lostaglio-coherence,Lostaglio-bounds, Cwiklinski-limitations, Marvian-asymmetry, Imam-asymmetry}. 
In particular, we have clarified that our results do not contradict any previous result about the non-increasing properties of coherence or asymmetry under thermal (or more generally phase covariant) operations.
Quite the contrary, our results point at unnoticed subtleties arising when considering bipartite systems with a global degenerate Hamiltonian, allowing e.g. the local amplification of coherence in both subsystems. 
These degeneracies are not particular of the specific setup we considered in this paper, but they are ubiquitous in quantum continuous devices acting as heat engines or refrigerators, either autonomous or non-autonomous \cite{Kosloff-review, Linden-fridge, Brunner-virtual, Silva-performance}.
Last, but not least, given the broad scope of asymmetry theories and their many applications to diverse problems in physics \cite{Marvian14}, it may be interesting to extend our results to more general symmetries others than time-translation.   
}

\begin{acknowledgments}
We thank Paul Skrzypczyk for interesting discussions {\blue and Imam Marvian for providing useful comments which helped us to improve this paper}. 
We acknowledge financial support from Spanish Government, grants TerMic (FIS2014-52486-R) and CONTRACT (FIS2017-83709-R), the European project ERC-AD NLST, and the Swiss National Science Foundation (grant No. PP00P2\_138917 and grant No. 200020 165843 and QSIT). G. M. acknowledges support from MINECO FPI grant No. BES-2012-054025. 
This work has been supported by the COST Action MP1209 ``Thermodynamics in the quantum regime''. 
\end{acknowledgments}

\appendix

\section{Asymmetry and coherence in bipartite systems}
\label{appC}

{\blue As mentioned in Sec. \ref{sec:thermo}, two measures of coherence based on relative entropy have been proposed in the literature: REC, $C(\rho)$ in Eq. \eqref{eq:relcoherence}, defined with respect to  an orthogonal basis $\mathcal B$ of the Hilbert space, and REA, $A(\rho)$ in Eq. \eqref{eq:relasymmetrt}, defined with respect to an operator, usually the Hamiltonian $H$.} 
For bipartite systems we can further elaborate upon the differences between these two quantities. Let $\rho$  be the state of a bipartite system $A+B$ with reduced states $\rho_A=\tr_B (\rho)$ and $\rho_B=\tr_A (\rho)$. Using the mutual information $I(\rho)=S(\rho_A)+S(\rho_B)-S(\rho)$, one can write
\begin{align}
C(\rho) & = C(\rho_{A})+C(\rho_{B})-I(\bar \rho)+I(\rho)\label{caddit}  \\
A(\rho) & = A(\rho_{A})+A(\rho_{B})-I(\tilde \rho)+I(\rho).\label{saddit}
\end{align}
If the constituents of the  bipartite system  are non-degenerate, the coherence and asymmetry of the reduced states are equal, $C(\rho_A)=A(\rho_A)$ and
$C(\rho_B)=A(\rho_B)$. In that case, 
the difference between coherence and asymmetry can be written as
\begin{equation}\label{c-a}
C(\rho)-A(\rho)=I(\tilde \rho)-I(\bar \rho)\geq 0.
\end{equation}

\subsection{Sub-additivity of asymmetry: an example}
\label{appC2}

Applying the previous relations (\ref{caddit}) and (\ref{saddit})  to an uncorrelated state $\rho=\rho_A\otimes \rho_B$ with $I(\rho)=0$, and assuming that the basis $\mathcal B$ is local, we get
\begin{align}
C(\rho) &= C(\rho_{A})+C(\rho_{B}) \nonumber \\
A(\rho)&-[A(\rho_{A})+A(\rho_{B})]=-I(\tilde \rho)\leq 0
\end{align}
since the fully dephased state is also uncorrelated with $I(\bar \rho)=0$. On the contrary,  partial dephasing can create spurious correlations between the two systems, $I(\tilde \rho)>0$, hence the asymmetry can decrease when considering the reduced states separately. This means that coherence is additive but asymmetry is sub-additive: 
\begin{equation}
A(\rho)\leq A(\rho_A)+A(\rho_B). 
\end{equation}

A simple example is given by two qubits, $A$ and $B$, with Hamiltonian $H=\epsilon\left[\ket{1}_A\bra{1}_A+\ket{1}_B\bra{1}_B\right]$, which is degenerate since states $\ket{01}$ and $\ket{10}$ have the same energy $\epsilon$. 
Here we discuss partially dephased states $\tilde \rho$ with respect to the eigen-projectors of the Hamiltonian $H$ and fully dephased states $\bar \rho$ with respect to the basis ${\mathcal B}=\{\ket{00},\ket{01},\ket{10},\ket{11}\}$, which is the only local eigenbasis of $H$.

Consider for instance  the pure states $\rho_A=\rho_B=\ket{\psi}\bra{\psi}$ with $\ket{\psi}=[\ket{0}+\ket{1}]/\sqrt{2}$. 
In matrix form, using the canonical local basis $\{\ket{0},\ket{1}\}$ and the global basis $\mathcal B$:
\begin{equation}
\rho=\frac{1}{2}\left(\begin{array}{cc}1 & 1 \\ 1 & 1 \end{array}\right) \otimes
\frac{1}{2}
\left(\begin{array}{cc} 1  & 1 \\ 1 & 1 \end{array}\right)=\frac{1}{4}\left(\begin{array}{cccc}
1 & 1 & 1 & 1 \\
1 & 1 & 1 & 1 \\
1 & 1 & 1 & 1 \\
1 & 1 & 1 & 1 
\end{array}\right).
\end{equation}
This global state is also a pure uncorrelated state, that is $I(\rho)=0$, but the partially dephased state
\begin{equation}
\tilde \rho=\frac{1}{4}\left(\begin{array}{cccc}
1 & 0 & 0 & 0 \\
0 & 1 & 1 & 0 \\
0 & 1 & 1 & 0 \\
0 & 0 & 0 & 1 
\end{array}\right) 
\end{equation}
exhibits correlations, namely, $I(\tilde \rho)=2\ln 2 - (\ln 4 + \ln 2)/2 = \ln 2 /2$. 

The coherence of  the reduced states is $C(\rho_A)=C(\rho_B)=S(\tilde \rho_{A})-S(\rho_A)=\ln 2$, and the coherence of the global state satisfies
\begin{equation}
C(\rho)=S(\bar \rho)-S(\rho)=2\ln 2 = C(\rho_A)+C(\rho_B).
\end{equation}
On the other hand, the asymmetry of the reduced states is still $A(\rho_A)=A(\rho_B)=\ln 2$, but the asymmetry of the global state is smaller than the sum of local coherences:
\begin{align}
A(\rho)&=S(\tilde \rho)-S(\rho)=\frac{3}{2}\ln 2 \nonumber \\ 
       &\leq 2\ln 2 = A(\rho_A)+A(\rho_B)
\end{align}
that is,  asymmetry is sub-additive. The difference between the asymmetry of the global state $A(\rho)=3\ln 2 /2$ and the sum of asymmetries of the reduced states $A(\rho_A)+A(\rho_B)=2\ln 2$ is precisely $I(\tilde \rho)=\ln 2/ 2$.

\subsection{Increase of coherence under thermal operations}
\label{appC3}

{\blue These results indicate that local asymmetries can indeed increase under (phase covariant) thermal operations when there exist degeneracies in the total Hamiltonian. Furthermore, as a complementary view, we may be also interested on the behavior of the global REC, $C(\rho)$, in such degenerate situations. 
Our analysis shows that for degenerate systems, the global REC can grow under thermal operations. These issues has been in fact largely overlook, since most research has focused on non-degenerate systems, or on REC within degenerate subspaces. For instance, many parts of the analysis in \cite{Lostaglio-bounds}, 
where they apply general results for the so-called ``modes of asymmetry'' \cite{Marvian-modes} to the study of thermal operations [see e.g. Eq.~(8) in that reference], do not hold if the global Hamiltonian present degeneracies.}

\begin{figure}
	\bigskip
	\flushright
	\includegraphics[width=1.0 \linewidth]{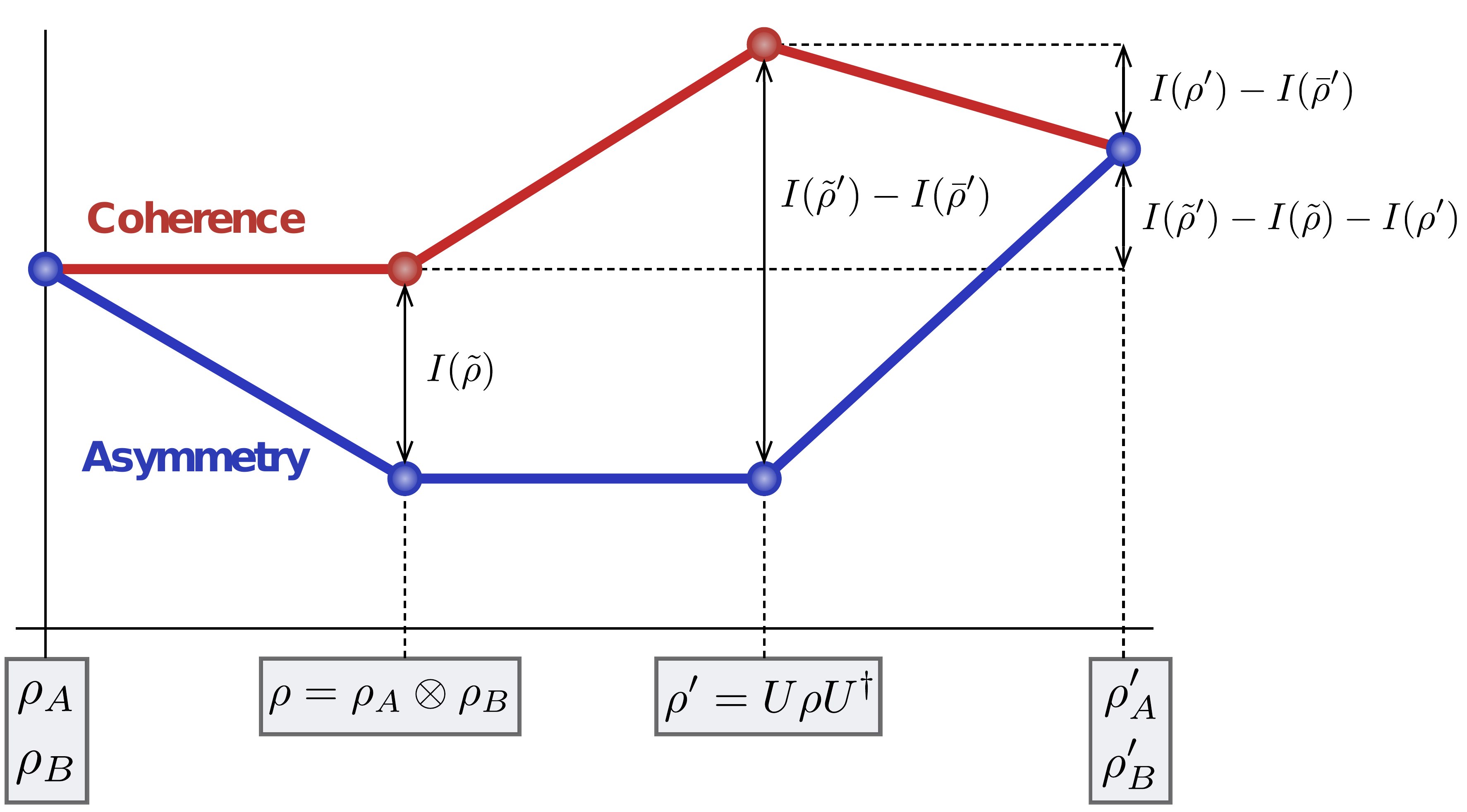}
	\caption{REC $C(\rho)$ (red) and REA $A(\rho)$ (blue) along the evolution of a bipartite system $A+B$, starting from an uncorrelated system $\rho=\rho_A\otimes \rho_B$. In the left and right points we plot $C(\rho_A) + C(\rho_B)$ and $C(\rho_A^\prime) + C(\rho_B^\prime)$ respectively. \label{fig:coherence_vrs_asym}}
\end{figure}

To discuss the above points in more detail, let us consider our bipartite system starting from an uncorrelated state $\rho=\rho_A\otimes\rho_B$ and evolving to $\rho'=U\rho U^\dagger$ according to a unitary operator $U$ that commutes with the global Hamiltonian $H$. Such a unitary transformation is a special case of a {\em thermal operation}, and as such, it is known to conserve asymmetry \cite{Lostaglio-bounds}. However, it is not hard to see that REC can increase under this type of operations. The evolution of coherence and asymmetry is sketched in Figure \ref{fig:coherence_vrs_asym}. We distinguish four stages in the process, depicted along the horizontal axis. First we consider the sum of local REC (REA) of the reduced states, $\rho_A$ and $\rho_B$. 
Second, we plot the REC (REA) of the global initial state $\rho=\rho_A\otimes\rho_B$. Even though this state is uncorrelated, the REA could experience a decrease $I(\tilde \rho)$  from stage 1 to 2 due to sub-additivity. In the third stage we compute the REC (REA) of the global state after the transformation $\rho'=U\rho U^{\dagger}$. The REA is conserved  but, according to (\ref{c-a}), the REC is $A(\rho')+I(\tilde \rho')-I(\bar \rho')$. Then the change of the global REC  due to the unitary transformation is 
$I(\tilde \rho')-I(\bar \rho')-I(\tilde \rho)$,
which can be positive, as we show below in an explicit example. {\blue The fourth stage is the result of ``separating'' the two qubits, that is, considering them as independent entities not allowed to interact again, and calculating the sum of local REC (REA) of each reduced state, $\rho'_A$ and $\rho'_B$}.  Notice that the local REC and local REA coincide, since the two systems $A$ and $B$ are non degenerate.

We see in the figure that REC always decreases when correlations are destroyed (neglected) $C(\rho)\geq C(\rho_A)+C(\rho_B)$ but it can increase under the unitary evolution $U$. On the other hand, REA is constant under evolution, but it can increase when the two systems are separated $A(\rho)\leq  A(\rho_A)+A(\rho_B)$. From the picture we conclude that REC can increase in the thermal operation if $I(\tilde \rho')>I(\bar \rho')+I(\tilde \rho)$, whereas both coherence and asymmetry increase after the whole process (evolution + separation)  if $I(\tilde \rho')> I(\tilde \rho)+I(\rho')$. Notice that even though the basis $\mathcal B$ is local, $I(\bar \rho')$ can be different from zero due to classical correlations between the two qubits. Summarizing, the change along the whole process of both the local REA and local REC is:
\begin{equation}\label{applocals}
\Delta C_A+\Delta C_B=I(\tilde \rho')- I(\tilde \rho)-I(\rho')
\end{equation}
where $\Delta C_i=C(\rho'_i)-C(\rho_i)=A(\rho'_i)-A(\rho_i)$ for $i=A,B$.

As an explicit example, consider the following initial state $\rho=\rho_A \otimes \rho_B$ with
\begin{equation}\label{appinit}
\rho_A= \left(\begin{array}{cc}1/2 & c \\c & 1/2\end{array}\right)
 \qquad 
\rho_B=\left(\begin{array}{cc}0 & 0 \\0 & 1\end{array}\right),
\end{equation}
$c$ being a real number in the interval $c\in [-1/2,1/2]$ to ensure the positivity of $\rho_A$. We choose the following as a unitary thermal transformation 
\begin{equation}
U= \left(\begin{array}{cccc}
1 & 0 & 0 & 0 \\
0 & 1/\sqrt{2} & -i/\sqrt{2} & 0 \\
0 & -i/\sqrt{2} & 1/\sqrt{2} & 0 \\
0 & 0 & 0 & 1
\end{array}\right).\label{unit1}
\end{equation}
The initial global state in the basis $\{\ket{00},\ket{01},\ket{10},\ket{11}\}$ reads from Eq. \eqref{appinit}
\begin{equation}
 \rho = \left(\begin{array}{cccc}
0 & 0 & 0 & 0 \\
0 & 1/2 & 0 & c \\ 
0 & 0 & 0 & 0 \\
0 & c & 0 & 1/2
\end{array}\right) 
\end{equation}
and the final (global) state in the same basis is
\begin{equation}
\rho'= U\rho U^\dagger = \left(\begin{array}{cccc}
0 & 0 & 0 & 0 \\
0 & 1/4 & i/4 & c/\sqrt{2} \\ 
0 & -i/4 & 1/4 & -ic/\sqrt{2} \\
0 & c/\sqrt{2} & ic/\sqrt{2} & 1/2
\end{array}\right) 
\end{equation}
with final reduced states
\begin{equation}
 \rho'_A=\left(\begin{array}{cc}1/4 & c/\sqrt{2} \\c/\sqrt{2} & 3/4\end{array}\right) \qquad
\rho'_B=\left(\begin{array}{cc}1/4 & -ic/\sqrt{2} \\ ic/\sqrt{2} & 3/4\end{array}\right).
\end{equation}

\begin{figure}
	\bigskip
	\flushright
	\includegraphics[width=1.0 \linewidth]{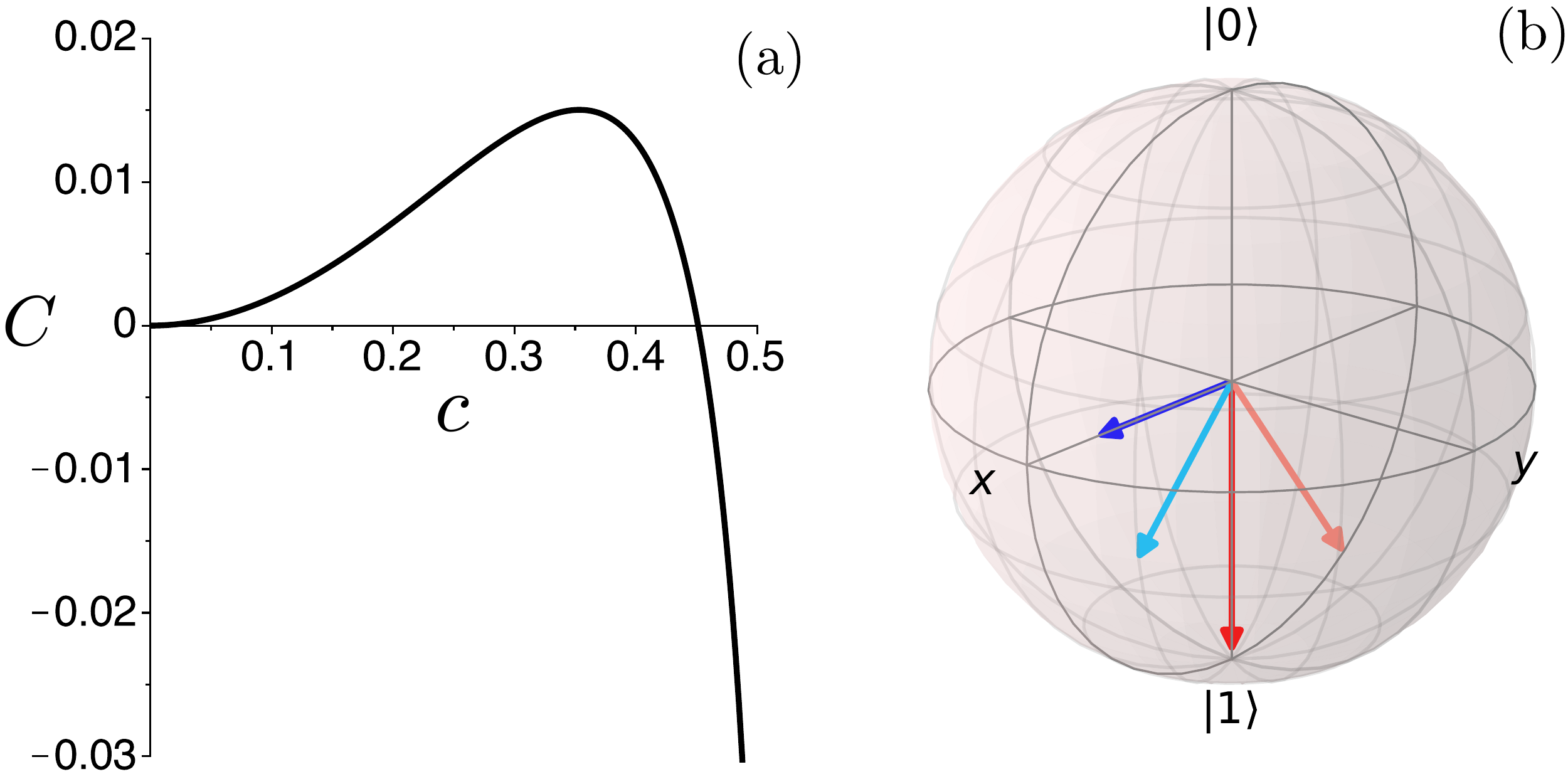}
	\caption{(a) Increase of REC, $\Delta C\equiv C(\rho'_A)+C(\rho'_B)-C(\rho_A)-C(\rho_B)$, as a function of $c$ for the example discussed in the text. (b) Evolution of the reduced states in the Bloch sphere for $c=0.35$. The initial states are $\rho_A=[2c,0,0]$ (dark blue) and $\rho_B=[0,0,-1]$ (red)   and the final ones $\rho'_A=[2c/\sqrt{2},0,-1/2]$ (light blue)  and  $\rho'_A=[0,2c/\sqrt{2},-1/2]$ (orange) \label{fig:deltacb}}
\end{figure}
\noindent
Notice that the partially dephased final state
\begin{equation}
\tilde \rho'=\left(\begin{array}{cccc}
0 & 0 & 0 & 0 \\
0 & 1/4 & i/4 & 0 \\ 
0 & -i/4 & 1/4 & 0  \\
0 & 0  & 0 & 1/2
\end{array}\right) 
\end{equation}
exhibits correlations between the two systems. The mutual information of the relevant states read
\begin{align}
I(\rho) &= I(\tilde \rho)=I(\bar \rho)=0,  \\
I(\tilde \rho')&= 2h(1/4) - h(1/2), \\   
I(\rho') &=  2h(p)- h(1/2+c), \\
I(\bar \rho') &= 2h(1/4) - 3 h(1/2)/2,
\end{align}
where $h(p)\equiv -p\ln p - (1-p)\ln(1-p)$ is the binary Shannon entropy and  $p=1/2 + \sqrt{1+8c^2}/4$.
The final increase of REC or REA is (see Fig.~\ref{fig:coherence_vrs_asym})
\begin{align}
\Delta C & \equiv  C(\rho'_A)+C(\rho'_B)-[C(\rho_A)+C(\rho_B)]\nonumber \\
&= I(\tilde \rho')-I(\tilde \rho)-I(\rho')\nonumber \\
&= 2\,h(1/4) - h(1/2)-2\,h(p)+ h(1/2+c).
\end{align}
This increase of REC is shown in Fig.~\ref{fig:deltacb}(a) as a function of $c$. 
There we can see that the total REC increases for a wide range of the parameter $c\leq 0.4513\dots$ 

It is illustrative to see the evolution of the two reduced states in the Bloch sphere, as shown in Fig.~\ref{fig:deltacb}(b) for $c=0.35$. 
This figure partly illustrates the mechanism of our machine, although the specific initial condition (\ref{appinit}) is in general different from the states of the TLA and the machine in our setup. 
If we identify system $A$ with the machine and system $B$ with the atom, we see that the interaction is able not only to transfer local coherence from the machine to the atom, but also to increase the global REC. 
This is what happens when the atom interacts with the virtual qubit of the machine. 
The role of the thermal baths is to restore the machine virtual qubit to its initial value. 
However, the thermal baths cannot increase the coherence of the machine, as would be needed in this specific example (compare the initial and final states of the machine $\rho_A$ and $\rho'_A$). 
For the machine to work, it would be necessary to increase simultaneously the local coherences of the atom and the machine using a thermal operation. 
In the next section of this appendix we show that this is possible by generalizing the previous example.

\subsection{Simultaneous increase of local coherences}
\label{appC4}

Here we show that the simultaneous increase of the two local coherences is possible if one of the two qubits starts in a state with population inversion, i.e., with a higher probability to be in the excited state than in the ground state. To do so, let us generalize the last example by using the following family of thermal unitary operations:
\begin{equation}
U =e^{-i H_{\mathrm{int}}t/\hbar}= \left(\begin{array}{cccc}
1 & 0 & 0 & 0 \\
0 & \cos (\theta) & -i\sin (\theta) & 0 \\
0 & -i\sin (\theta) & \cos (\theta) & 0 \\
0 & 0 & 0 & 1
\end{array}\right),\label{unittheta}
\end{equation}
which are generated by the energy-preserving interaction Hamiltonian
\begin{equation} \label{eq:intapp}
H_{\mathrm{int}} = \hbar \omega \left( \ket{0}_A\ket{1}_{B} \bra{1}_A \bra{0}_B + \mathrm{h.c.} \right),
\end{equation}
and we have set $\theta=\omega t$. Notice that (\ref{unittheta}) for $\theta=\pi/4$ yields the transformation (\ref{unit1}) of our previous example, {\blue and the similarity of Eq. \eqref{eq:intapp} with the machine-atoms interaction Hamiltonian in Eq. \eqref{eq:Hint}}. 
We consider a general uncorrelated state  $\rho=\rho_A\otimes \rho_B$  as the initial condition, with reduced states 
\begin{equation}
\rho_A = \left(\begin{array}{cc}
\frac{1+\delta_A}{2} & c_A \\
c^*_A & \frac{1-\delta_A}{2} 
\end{array}\right) 
\qquad
\rho_B = \left(\begin{array}{cc}
\frac{1+\delta_B}{2} & c_B \\
c^*_B & \frac{1-\delta_B}{2} 
\end{array}\right).
\end{equation}
In this notation $-1 \leq \delta_i\leq 1$ denotes the \emph{bias} of system $i=A,B$, i.e., the difference between the populations of the ground and the excited states. 
The off-diagonal terms $c_i$ are complex numbers obeying $\delta_i^2+4|c_i|^2\leq 1$ to ensure the positivity of the density matrices. 
The REC or REA (the local Hamiltonians are non-degenerate) is $C(\rho_i)=h[(1+\delta_i)/2]-h[ (1+\sqrt{\delta_i^2+4|c_i|^2})/2]$, which is an increasing function of $|c_i|$.

The off-diagonal terms of the final reduced states can be expressed in a rather compact form:
\begin{align}
c_A^\prime &= c_A \cos(\theta) + i c_B \delta_A \sin(\theta) \\
c_B^\prime &= c_B \cos(\theta) + i c_A \delta_B \sin(\theta).
\end{align}
Now we can discuss the necessary conditions to achieve a simultaneous increase of both local coherences (or asymmetries). For that to occur, the following ratios must be larger than one:
\begin{align}
\frac{\left| c_A^\prime \right|}{\left| c_A \right|} &= \left|\cos(\theta) + i\,\delta_A\alpha\,e^{i\varphi}\sin(\theta)\right|> 1
\\
\frac{\left| c_B^\prime \right|}{\left| c_B \right|} &= \left|\cos(\theta) + i\frac{\delta_B}{\alpha}\,e^{-i\varphi}\sin(\theta)\right|> 1,
\end{align}
where we have introduced the modulus $\alpha$ and phase $\varphi$ of the ratio between the initial coherences: $c_B=\alpha e^{i\varphi}c_A$. 
The above inequalities can then be written as
\begin{align}
\cos^2(\theta)+\delta_A^2\alpha^2\sin^2(\theta) -2\delta_A\alpha\sin(\varphi) \sin(\theta)\cos(\theta)
 &>  1\label{cond1}
\\
\cos^2(\theta)+\frac{\delta_B^2}{\alpha^2}\sin^2(\theta) +\frac{2\delta_B}{\alpha}\sin(\varphi) \sin(\theta)\cos(\theta)
 &>  1 \label{cond2}
\end{align}
which, after some algebra and for $\sin(\theta) \neq 0$, reduce to
\begin{align}
\left(\delta_A\alpha -\kappa\right)\left(\delta_A\alpha +\frac{1}{\kappa}\right) &> 0 \nonumber \\
\left(\frac{\delta_B}{\alpha} -\frac{1}{\kappa}\right)\left(\frac{\delta_B}{\alpha} +\kappa\right) &> 0
\end{align}
with $\kappa=(r+\sqrt{r^2+1})/2\geq 0$ and $r=\sin(\varphi) \cot(\theta)$. These two inequalities imply that the biases $\delta_A$ and $\delta_B$ have opposite signs. To prove this, suppose that both are positive. In this case
\begin{equation}
\delta_A\alpha >\kappa,\qquad \frac{\delta_B}{\alpha} >\frac{1}{\kappa}
\end{equation}
and multiplying both inequalities one gets $\delta_A\delta_B>1$, which is not possible since the biases are bound between $-1$ and 1. The case when both $\delta_A$ and $\delta_B$ are negative is analogous.

\begin{figure}
	\bigskip
	\centering
	\includegraphics[width=0.95 \linewidth]{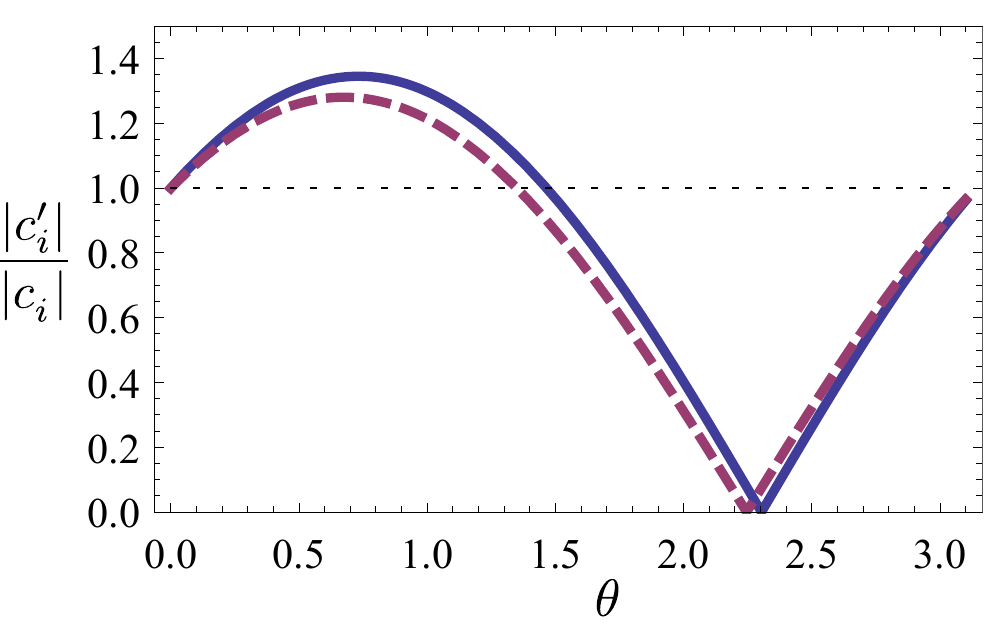}
	\caption{Simultaneous amplification of local coherences via a thermal operation on a two qubit system. The operation is the unitary transformation (\ref{unittheta}). The figure shows the ratio $|c'_i|/|c_i|$  between the modulus of the off-diagonal terms in the density matrices of the two qubits, $i=A$ (solid), $B$ (dashed), after and before the transformation, as a function of the parameter $\theta$. The horizontal dotted line at 1 is shown for reference. The rest of the parameters are $\delta_A=-0.9$, $\delta_B=0.8$, $\alpha=1$, and $\varphi=\pi/2$, that is $c_B=ic_A$.\label{fig:cohplot}}
\end{figure}

If the biases have opposite sign, $\delta_A\delta_B<0$, it is always possible to find parameters for which the local coherences increase. An example is given in Fig.~\ref{fig:cohplot}, where we plot the ratios $|c'_i|/|c_i|$ ($i=A,B$) as a function of $\theta$, for  $\delta_A=-1$, $\delta_B=0.8$, $\alpha=1$, and $\varphi=\pi/2$, that is $c_B=ic_A$.

In summary, within this appendix we have derived a number of results that help to understand how our machine works and what is the role of each component. 
The interaction between the atom and the machine is similar to the unitary transformation (\ref{unittheta}), capable of enhancing the coherence of the two qubits, atom and machine, if there is a population inversion in one of the two systems and some coherence in the two initial reduced states. 
Consequently, if the TLA populations are not inverted, for a steady increase of coherence, it is necessary to restore the machine to a state with some coherence and population inversion. There the two thermal baths come into play. The difference of temperature of the baths creates the required population inversion in the virtual qubit. 
Finally, to restore the machine to a state with coherence, we must prevent it from fully relaxing to the steady state under the influence of the two baths. Summarizing, a partial thermal relaxation  in contact with the two baths at different temperatures  is capable of restoring the machine to a state with some remaining coherence and a population inversion. 
Then the unitary transformation can be repeated and induce a steady increase of coherence in the TLA.

\section{Derivation of the dynamics}\label{appA}

In order to derive master equations for the dynamical evolution of the thermal machine and the TLA stream we assume that the two-qubit machine is weakly coupled to thermal reservoirs modelled 
by a collection of bosonic modes $H_\alpha = \sum_k \hbar \Omega_k^{(\alpha)} b_k^{\alpha~ \dagger} b_k$ for $\alpha= 1, 2$ and where $[b_k^{(\alpha)}, b_{k^\prime}^{(\alpha^\prime)}] = 
\delta_{k, k^\prime} \delta_{\alpha, \alpha^\prime}$, in equilibrium Gibbs states. Their interaction in the rotating-wave approximation reads
\begin{equation}
 H_{\mathrm{int}} = \sum_{\alpha = 1,2} \sum_k \hbar g_k^{\alpha} \left(\sigma_\alpha^{~} b_k^{(\alpha) \dagger} + \sigma_\alpha^\dagger b_k^{\alpha} \right),
\end{equation}
where the parameters $g_k^{\alpha}$ control the coupling strength of the qubit $\alpha$ to each mode $k$ in the corresponding reservoir 
as specified by their spectral densities $J_\alpha(\Omega) = \sum_k \frac{(g_k^{\alpha})^2}{ \Omega_k^{\alpha}} \delta( \Omega - \Omega_k^\alpha)$. 

\subsection{Master equation for the machine} \label{appAmaster}

In the absence of the TLA stream and assuming Ohmic dissipation { within the standard Born-Markov and rotating-wave approximations}, the machine evolves in the interaction picture according to the following master equation 
in Lindblad form \cite{Bre02}
\begin{equation}
 \dot{\rho}_\mathrm{m} = \mathcal{L}_0 (\rho_\mathrm{m}) = \mathcal{D}_1(\rho_\mathrm{m}) + \mathcal{D}_2 (\rho_\mathrm{m}), 
\end{equation}
where we obtain two dissipators describing the exchange of energy quanta with each reservoir
\begin{align} \label{dissipator}
\mathcal{D}_\alpha(\rho_\mathrm{m}) &=~ k_{\downarrow}^{\alpha} \left( \sigma_\alpha \rho_\mathrm{m} \sigma_\alpha^\dagger - \frac{1}{2} \{ \sigma_\alpha^\dagger \sigma_\alpha, \rho_\mathrm{m} \} \right) \\
 & + k_{\uparrow}^{\alpha} \left( \sigma_\alpha^\dagger \rho_\mathrm{m} \sigma_\alpha - \frac{1}{2} \{ \sigma_\alpha \sigma_\alpha^\dagger, \rho_\mathrm{m} \} \right), ~~ \alpha= 1,2. \nonumber
\end{align}
In the above equation the rates $k_\downarrow^{\alpha} = \gamma_0^\alpha (n_\mathrm{th}^\alpha + 1)$ and $\gamma_\uparrow = \gamma_0^\alpha n_\mathrm{th}^\alpha$ depend on the mean number of thermal 
excitations in the reservoirs $n_\mathrm{th}^\alpha = (e^{\beta_\alpha E_\alpha} - 1)^{-1}$ and the spontaneous emission rates $\gamma_0^\alpha \ll E_\alpha^\prime~~ \forall \alpha, \alpha^\prime$.

We then model the interaction of the TLA stream with the dissipative two-qubits machine. Following the main text [Fig \ref{fig:sketch}(a)], the atoms interact one at a time with the machine for a short interval of time $\tau_\mathrm{i}$ 
according to the interaction Hamiltonian in Eqs. (\ref{eq:mastermachine}) and (\ref{eq:masteratoms}). This leads to the following unitary acting on the compound machine-atom system
\begin{align}\label{eq:unitary}
 U (t + \tau_\mathrm{i}, t) & = \exp \left( -\frac{i}{\hbar} \int_{t}^{t + \tau_\mathrm{i}} H_\mathrm{ma}(s) ds \right) \nonumber \\
 & = \exp \left( -i \phi V \right) \simeq \mathbb{I} - i \phi V - \frac{\phi^2}{2} V^2,
\end{align}
where we used $\phi \ll 1$ as defined in the main text, and $t$ is arbitrary. At this point we make a crucial assumption, namely, that the interaction time is short compared with the relevant timescales of the machine relaxation 
dynamics, $\tau_\mathrm{i} \ll 1/\gamma_0^{\alpha}$ for $\alpha = 1, 2$. In this case the state of the compound system during $\tau_\mathrm{i}$ evolves as
\begin{align} \label{eq:totaleffect}
 & \rho(t + \tau_\mathrm{i}) =  U (t + \tau_\mathrm{i}, t) \rho(t) U (t + \tau_\mathrm{i}, t)^\dagger   \\ 
 & = \rho(t) - i \phi [V, \rho(t)] + \phi^2 \left( V \rho(t) V - \frac{1}{2} \{V^2, \rho(t) \} \right), \nonumber 
\end{align}
that is, we neglect the action of the thermal reservoirs during the interaction between the machine and the flying atom. Furthermore, we assumed that the machine and atom were initially uncorrelated, and the machine always interacts with a `fresh' atom prepared in the same initial state $\rho(t) = \rho_\mathrm{m}(t) \otimes \rho_\mathrm{a}$.

Let us denote the effective action of a single TLA on the machine as the completely-positive and trace-preserving (CPTP) map $\mathcal{E}(\rho_\mathrm{m}) = \tr_\mathrm{a}[\rho(t + \tau_\mathrm{i})]$. 
The evolution of the machine at some time $t$ after $n$ interactions can be then written as \cite{Espositonew}:
\begin{equation}
 \rho_\mathrm{m}^{(n)}(t) = \int_{t_0}^{t} ds ~w(t - s) e^{\mathcal{L}_0 (t - s)} \mathcal{E}(\rho_\mathrm{m}^{(n-1)}(s)),
\end{equation}
where $w(t)$ is the waiting time distribution, which characterizes how much time we need to wait from one interaction to the next. We assume Poisson statistics $w(t) = r e^{-r t}$, where $r$ is the average 
rate at which interactions occur. Now taking the time-derivative of the above equation, and summing over $n$ (see Ref. \cite{Espositonew} for more details), we obtain the master equation (\ref{eq:mastermachine}):
\begin{equation} \label{eq:mastermachineapp}
\dot{\rho}_\mathrm{m} = -i r \phi [V_\mathrm{m}, \rho_\mathrm{m}] + \mathcal{D}_a(\rho_\mathrm{m}) + \mathcal{L}_0 (\rho_\mathrm{m}) \equiv \mathcal{L}_\mathrm{m}(\rho_\mathrm{m}), 
\end{equation}
where we obtained a new dissipator reading
\begin{align} \label{eq:dissmachine}
 \mathcal{D}_\mathrm{v}(\rho_\mathrm{m}) &=~ r \phi^2 \langle \sigma_\mathrm{a}^{~} \sigma_\mathrm{a}^\dagger \rangle_0 \left( \sigma_\mathrm{v} \rho_\mathrm{m} \sigma_\mathrm{v}^\dagger - \frac{1}{2} \{ \sigma_\mathrm{v}^\dagger \sigma_\mathrm{v}, \rho_\mathrm{m} \} \right)  \nonumber \\ 
 & + r \phi^2 \langle \sigma_\mathrm{a}^\dagger \sigma_\mathrm{a}^{~} \rangle_0 \left( \sigma_\mathrm{v}^\dagger \rho_\mathrm{m} \sigma_\mathrm{v} - \frac{1}{2} \{ \sigma_\mathrm{v} \sigma_\mathrm{v}^\dagger, \rho_\mathrm{m} \} \right). 
\end{align}
We recall that here $V_\mathrm{m}= \sigma_\mathrm{v} \langle \sigma_\mathrm{a}^\dagger \rangle + \sigma_\mathrm{v}^\dagger \langle \sigma_\mathrm{a} \rangle$, $\sigma_\mathrm{v} = \sigma_1^\dagger \sigma_2^{~}$ being the lowering operator of the virtual qubit of the machine, and the expectation value $\langle \sigma_\mathrm{a}^{~} \sigma_\mathrm{a}^\dagger \rangle = \tr_\mathrm{a}[\sigma_\mathrm{a}^{~} \sigma_\mathrm{a}^\dagger \rho_\mathrm{a}]$ is the initial probability to find the TLA in its ground state. Analogously, the term $\langle \sigma_\mathrm{a}^\dagger \rangle = \tr_\mathrm{a}[\sigma_\mathrm{a}^\dagger \rho_\mathrm{a}]$ represents the initial coherence in the atoms. Notice that the coherent term in Eq. (\ref{eq:mastermachineapp}) will acquire a time-dependent modulation when turning back to the Schr\"odinger picture, so that one must keep trace of its phase during the evolution in practical applications.

It is worth mentioning that in our derivation of the machine dynamics, our assumptions naturally agree with the local approach for modeling Lindblad master equations \cite{Hofer,Onam}. This is because the qubits of the machine do not interact 
between them most of the time, but only with their respective reservoirs. The only interaction between them is indeed during the time in which a TLA passes by the machine, which is assumed to be small ($\tau_\mathrm{i} \ll 1/\gamma_0^\alpha$ for 
$\alpha = 1, 2$). One may consider longer timescales of interaction for the atoms, and in that case compare the local and global approaches. This is an interesting question, but outside the focus of the present work. We expect that the extra dissipation 
channels that arise in the global approach would produce undesirable heat flows reducing the power and performance of the machine, such as has been pointed out in Ref. \cite{Correa-leaks}.

\subsection{Two-level atoms CPTP map}

The state change of any flying TLA due to its interaction with the machine $\rho_\mathrm{a} \rightarrow \rho_\mathrm{a}^\prime$ can be also obtained from this model. We denote the effective action of the machine in the TLA as the CPTP 
map $\mathcal{A}_t(\rho_\mathrm{a}) = \tr_\mathrm{m}[\rho(t + \tau_\mathrm{i})]$ for $\rho(t + \tau_\mathrm{i})$ given in Eq. (\ref{eq:totaleffect}). We obtain:
\begin{equation} \label{eq:atomsmap}
\mathcal{A}_t(\rho_\mathrm{a}) = \rho_\mathrm{a} -i \phi [ V_\mathrm{a}, \rho_\mathrm{a}]~ + ~\mathcal{D}_{\mathrm a}(\rho_\mathrm{a})
\end{equation}
where $V_\mathrm{a} = \sigma_\mathrm{a} \langle \sigma_\mathrm{v}^\dagger \rangle_t + \sigma_\mathrm{a}^\dagger \langle \sigma_\mathrm{v} \rangle_t$, and we obtain the dissipator complementary 
to (\ref{eq:dissmachine})
\begin{align} \label{eq:dissatom}
 \mathcal{D}_\mathrm{a}(\rho_\mathrm{a}) &=~ \phi^2 \langle \sigma_\mathrm{v}^{~} \sigma_\mathrm{v}^\dagger \rangle_t \left( \sigma_\mathrm{a} \rho_\mathrm{a} \sigma_\mathrm{a}^\dagger - \frac{1}{2} \{ \sigma_\mathrm{a}^\dagger \sigma_\mathrm{a}, \rho_\mathrm{a} \} \right) \nonumber \\
 & + \phi^2 \langle \sigma_\mathrm{v}^\dagger \sigma_\mathrm{v}^{~} \rangle_t \left( \sigma_\mathrm{a}^\dagger \rho_\mathrm{a} \sigma_\mathrm{a} - \frac{1}{2} \{ \sigma_\mathrm{a} \sigma_\mathrm{a}^\dagger, \rho_\mathrm{a} \} \right). 
\end{align}
Notice that this dissipator does not depend on $r$, as the state change in any atom in the sequence is independent of the rate at which atoms are 
sent through the machine. Furthermore the expectation values are time-dependent, that is $\langle \sigma_\mathrm{v}^\dagger \rangle_t = \tr_\mathrm{m}[\sigma_\mathrm{v}^\dagger \rho_\mathrm{m}(t)]$ and analogously for 
$\langle \sigma_\mathrm{v}^\dagger \sigma_\mathrm{v}^{~} \rangle_t$ and $\langle \sigma_\mathrm{v}^{~} \sigma_\mathrm{v}^{\dagger} \rangle_t$, coming from the fact that the change in the state of any atom in the sequence 
depends on the actual state of the machine. Henceforth we have a CPTP map $\mathcal{A}_t(\rho_\mathrm{a})$ for any given state of the machine $\rho_\mathrm{m}(t)$, that is, for any given instant of time $t$.
It is only when the two-qubit machine reaches a steady state, that it will produce the same time-independent kick $\mathcal{A}(\rho_\mathrm{a})$ on input atoms arriving in the same initial sate 
$\rho_\mathrm{a}$. Under these conditions, $\mathcal{A}(\rho_\mathrm{a})$, represents the average state of all output atoms.

\subsection{Extended confituration} \label{appAext}

Finally, we consider the configuration presented in Fig. \ref{fig:sketch2}. In this case we have a large sequence of two-qubit machines into which input atoms prepared in $\rho_\mathrm{a}^0$ are sent.
Therefore the first machine in the sequence is just described by our above reasoning. Moreover, we can extend the argument to each machine in the sequence by simply replacing the initial state in which the atoms are prepared $\rho_\mathrm{a}^0$, by some arbitrary state $\rho_\mathrm{a}$ representing the state of the TLA at the beginning of the interaction with any machine. 
This state will of course depend on the previous interaction of the atom with the preceding machines in the sequence and will be therefore different for each of them. Accordingly, each machine will now produce a different kick on the TLA state [Eqs. (\ref{eq:atomsmap}) and (\ref{eq:dissatom})], depending on its time-dependent state, which in turn depends on the previous atoms which have already interacted with it. This complicated situation is however greatly simplified in the case in which all the machines in the sequence may reach a steady state (as we argue in the main text). In that case each machine still produces a different kick $\mathcal{A}_i(\rho_\mathrm{a})$ as 
its state depends on its position $i$ in the sequence, but following Eqs. (\ref{eq:mastermachineapp}) and (\ref{eq:dissmachine}), this state will only depend on the state of their input atoms $\rho_\mathrm{a}$, leading to Eq. (\ref{eq:mapstream}).

{\blue Nevertheless, we notice that the above reasoning only leads to the dynamical evolution in Eq. \eqref{eq:mapstream} in the case in which the machines in the sequence get not correlated between them. In the following we provide a general argument to justify the validity of such assumption.
The generation of such correlations may only occur via consecutive interactions of any TLA with subsequent machines, which could be then viewed as an effective way of interaction between them. 
Consequently, the rate $\Gamma$ at which such correlations can be generated depends on both the rate $r$ at which TLA are sent through the sequence of machines and, crucially, the distance between consecutive the machines. 
Notice that this generation of correlations will compete with the effect of the thermal reservoirs, which are constantly interacting with any machine in the array, inducing an exponential decay of correlations at a rate $\gamma_0^\alpha$, for $\alpha = 1,2$. 
Therefore, it suffices to ensure that the generation of correlations between machines is slower than this decay, that is, $\Gamma \ll \gamma_0^\alpha$. This is indeed the case by assuming that the machines are sufficiently far away from each others in the array, 
a condition that can be imposed without decreasing the rate $r$ at which atoms are sent. This implies that we can make $\Gamma$ arbitrarily small compared to $\gamma_0^\alpha$ and, therefore, safely consider that the machines in the array are not able to correlate between them. 
Finally, notice that this is in contrast with the behavior of the correlations generated between consecutive TLA in the sequence, which depend only on $r$, and may occur even for a single machine.}

\section{Operation at steady state conditions}\label{appB}

As pointed out in the main text, our machine is able to operate in the steady state regime, that is, when sufficiently many TLAs have already interacted with it.
The steady state of the two-qubit machine $\pi_\mathrm{m}$ can be analytically obtained from the master equation (\ref{eq:mastermachine}) by imposing 
$\mathcal{L}_\mathrm{m}(\pi_\mathrm{m}) = 0$, which leads to:
\begin{align}
\pi_\mathrm{m} =&~~ \pi_{00} \ket{0}_1 \ket{0}_2 \bra{0}_1 \bra{0}_2 + \pi_{10} \ket{1}_1 \ket{0}_2 \bra{1}_1 \bra{0}_2 \nonumber \\ 
&+ \pi_{01} \ket{0}_1 \ket{1}_2 \bra{0}_1 \bra{1}_2 + \pi_{11} \ket{1}_1 \ket{1}_2 \bra{1}_1 \bra{1}_2  \nonumber \\
&+ \pi_\mathrm{v} \ket{0}_1 \ket{1}_2 \bra{1}_1 \bra{0}_2 + \pi_\mathrm{v}^\ast \ket{1}_1 \ket{0}_2 \bra{0}_1 \bra{1}_2.
\end{align}
Here $\pi_{00} + \pi_{01} + \pi_{10} + \pi_{11} = 1$ are the steady state populations of the four levels of the machine, and $\pi_\mathrm{v} = \tr[\sigma_\mathrm{v} \pi_\mathrm{m}]$ is 
the steady state coherence in the virtual qubit subspace. Recall that in the main text we have introduced the notation $\{ \ket{0}_\mathrm{v} \equiv \ket{1}_1 \ket{0}_2$, 
$\ket{1}_\mathrm{v} \equiv \ket{0}_1 \ket{1}_2 \}$ for the virtual qubit energy levels, together with the lowering operator $\sigma_\mathrm{v} \equiv \sigma_1^\dagger \sigma_2$. 
Once we substitute these values in the coefficients appearing in Eq. (\ref{eq:atomsmap}), the latter gives us the average output state of the TLA stream in the stationary regime.

We now focus on the values of the heat flows and energy currents: 
\begin{align}
\dot{Q}_k &\equiv  \tr[H_\mathrm{m} \mathcal{D}_k(\rho)], ~~~ k=1,2, \\ \label{eq:atomenergy}
\dot{E}_\mathrm{a} & \equiv r \tr[H_\mathrm{a} \big( \mathcal{A}_t(\rho_\mathrm{a}) - \rho_\mathrm{a} \big) ] \nonumber \\
&= \tr[H_\mathrm{m} \left( -i r \phi [V_\mathrm{m}, \rho_\mathrm{m}] + \mathcal{D}_\mathrm{m}(\rho_\mathrm{m}) \right)].
\end{align}
Here we have reintroduced $r$ in Eq.~(\ref{eq:atomenergy}) to calculate the rate at which energy is transferred to output atoms and the last equality 
follows as a consequence of the energy-preserving interaction between machine and atoms. 
In the steady state regime we obtain
\begin{align} \label{eq:steadyflows}
& \dot{Q}_2  = E_2 (\Delta_\mathrm{p} + \zeta_\mathrm{c}),~~~~  \dot{Q}_1  = - E_1 (\Delta_\mathrm{p} + \zeta_\mathrm{c}), \nonumber \\ 
& \dot{E}_\mathrm{a} = (E_2 - E_1) (\Delta_\mathrm{p} + \zeta_\mathrm{c}),  
\end{align}
where we introduced the key quantities:
\begin{align} \label{eq:parameters}
\Delta_\mathrm{p} & \equiv~ r \phi^2 (\pi_{01} \langle \sigma_\mathrm{a}^{~} \sigma_\mathrm{a}^\dagger \rangle  - \pi_{10} \langle \sigma_\mathrm{a}^{\dagger} \sigma_\mathrm{a}^{~} \rangle), \nonumber \\
\zeta_\mathrm{c} & \equiv~  i r \phi (\pi_\mathrm{v}^\ast \langle \sigma_\mathrm{a} \rangle- \pi_\mathrm{v}^{~} \langle \sigma_\mathrm{a}^\dagger \rangle).
\end{align}
Here, the atom averages are taken over  $\rho_\mathrm{a}$ and $\Delta_\mathrm{p}$ can be interpreted as the relative bias between the populations of the virtual qubit in the steady state and the TLA populations, which fulfills $\Delta_\mathrm{p} \geq 0 \Leftrightarrow \pi_{01}/\pi_{10} \geq \langle \sigma_\mathrm{a}^{\dagger} \sigma_\mathrm{a}^{~} \rangle/ \langle \sigma_\mathrm{a}^{~} \sigma_\mathrm{a}^{\dagger} \rangle$, i.e. it is positive only when the virtual qubit has a larger population inversion than the initial state of the TLA. On the other hand, the real number $\zeta_\mathrm{c}$ is always positive $\zeta_\mathrm{c} \geq 0$, and proportional to the square modulus of the initial coherence of the TLA $|\langle \sigma_\mathrm{a} \rangle|^2$. From Eq. (\ref{eq:steadyflows}) it is now easy to check 
that the following proportionality relation holds 
\begin{equation}
\frac{\dot{E}_\mathrm{a}}{E_2 - E_1} = \frac{\dot{Q}_2}{E_2} = -\frac{\dot{Q}_1}{E_1}.
\end{equation}
This relation has been demonstrated for the original model of the two-qubit machine we employ here \cite{Brunner-virtual}, being a consequence of the fact that each energy flow through the machine is 
mediated by a single transition. 

Finally, for computing free energy and coherence flows we need to calculate the average change in the von Neumann entropy of the TLA stream in steady state conditions: 
\begin{equation}
 \dot{S}_\mathrm{a} \equiv r \big[ -\mathcal{A}(\rho_\mathrm{a})\ln \mathcal{A}(\rho_\mathrm{a}) + \rho_\mathrm{a} \ln \rho_\mathrm{a} \big].
\end{equation}
This can be done by applying perturbation theory to calculate the eigenvalues and eigenstates of $\mathcal{A}(\rho_\mathrm{a}) \ket{\lambda_n} = \lambda_n \ket{\lambda_n}$. 
We expand $\lambda_n$ and $\ket{\lambda_n}$ up to second order in $\phi$, and identify the corresponding contributions in Eq. (\ref{eq:atomsmap}). The entropy change of the TLA 
stream can be calculated in this way as:
\begin{align} \label{eq:entropycalculus}
 \dot{S}_\mathrm{a} &= - r \sum_n \lambda_n \ln \lambda_n + r\sum_n \lambda_n^{(0)} \ln \lambda_n^{(0)}  \nonumber \\
 & \simeq - r \phi^2 \sum_n \lambda_n^{(2)} \ln \lambda_n^{(0)},
\end{align}
where $\lambda_n^{(2)}$ is the second-order contribution to the eigenvalue expansion, $\lambda_n \simeq \lambda_n^{(0)} + \lambda_n^{(2)} \phi^2$ (as long as $\lambda_n^{(1)} = 0$)  and $\lambda_n^{(0)}$ is the 
zeroth-order one, that is $\rho_\mathrm{a} \ket{\lambda_n^{(0)}} = \lambda_n^{(0)} \ket{\lambda_n^{(0)}}$. Therefore we just need to calculate $\lambda_n^{(2)}$. We obtain:
\begin{align} \label{eq:secondorder}
\lambda_n^{(2)} =& \phi^{-2} \bra{\lambda_n^{(0)}} \mathcal{D}_\mathrm{a}(\rho_\mathrm{a}) \ket{\lambda_n^{(0)}} \nonumber \\
& - \sum_{k \neq n} (\lambda_k^{(0)} - \lambda_n^{(0)})|\bra{\lambda_n^{(0)}} V_\mathrm{a} \ket{\lambda_k^{(0)}}|^2,
\end{align}
where the second term in the above equation comes from a non-zero first-order correction to the corresponding eigenstate, $\ket{ \lambda_n^{(1)}} =-i \sum_{k \neq n} \bra{\lambda_k^{(0)}} V_\mathrm{a} \ket{\lambda_n^{(0)}} \ket{\lambda_k^{(0)}}$.
Introducing Eq. (\ref{eq:secondorder}) into Eq. (\ref{eq:entropycalculus}) and operating, we finally arrive at
\begin{align} \label{eq:entropychange}
 \dot{S}_\mathrm{a} \simeq&~  \Big[\frac{\Delta_\mathrm{p}(\langle \sigma_\mathrm{a}^{\dagger} \sigma_\mathrm{a}^{~} \rangle - \langle \sigma_\mathrm{a}^{~} \sigma_\mathrm{a}^{\dagger} \rangle) - N_\mathrm{p} | \langle \sigma_\mathrm{a} \rangle|^2}{\lambda_{+}^{(0)} - \lambda_{-}^{(0)}} \nonumber \\ 
 & +~ r \phi^2|\pi_\mathrm{v}|^2 (\lambda_{+}^{(0)} - \lambda_{-}^{(0)})  \Big] \ln \left( \frac{\lambda_{-}^{(0)}}{\lambda_{+}^{(0)}} \right),   
\end{align}
where we have taken $\rho_\mathrm{a} = \rho_\mathrm{a}^0$ and introduced
\begin{align}\label{eq:eigenvalues}
 N_\mathrm{p} \equiv&~ r \phi^2 (\pi_{10} + \pi_{01}), \nonumber \\
 \lambda_{\pm}^{(0)} =&~ \frac{1}{2} \left( 1 \pm \sqrt{(\langle \sigma_\mathrm{a}^{\dagger} \sigma_\mathrm{a}^{~} \rangle- \langle \sigma_\mathrm{a}^{~} \sigma_\mathrm{a}^{\dagger} \rangle)^2 + 4 | \langle \sigma_\mathrm{a} \rangle|^2} \right),
\end{align}
the latter being the eigenvalues of $\rho_\mathrm{a}^0$. Eq. (\ref{eq:entropychange}) is to be compared with the entropy change in the state $\bar {\rho}_\mathrm{a}$ dephased in the $H_\mathrm{a}$ basis, which simply reads:
\begin{equation} \label{eq:dephasedentropychange}
 \dot{{S}}(\bar \rho_\mathrm{a}) = (\Delta_\mathrm{p} + \zeta_\mathrm{c})\ln \frac{\langle \sigma_\mathrm{a}^{~} \sigma_\mathrm{a}^{\dagger} \rangle}{\langle \sigma_\mathrm{a}^{\dagger} \sigma_\mathrm{a}^{~} \rangle}.
\end{equation}
The quantities in Eqs. (\ref{eq:steadyflows}), (\ref{eq:entropychange}) and (\ref{eq:dephasedentropychange}), together with the parameters in Eqs. (\ref{eq:parameters}) and (\ref{eq:eigenvalues}), 
are all we need to obtain all of the results presented in the main text.

\bibliography{references2}

\end{document}